\documentclass[twocolumn,superscriptaddress,preprintnumbers,amsmath,amssymb,prl]{revtex4}
\usepackage{graphicx}
\UseRawInputEncoding

\begin{document}

\thispagestyle{empty}

\title{Experimental and theoretical investigation of the thermal effect
in the Casimir interaction from graphene
}

\author{M. Liu}
\affiliation{Department of Physics and Astronomy, University of California, Riverside, California 92521, USA}

\author{Y. Zhang}
\affiliation{Department of Physics and Astronomy, University of California, Riverside, California 92521, USA}

\author{
G.~L.~Klimchitskaya}
\affiliation{Central Astronomical Observatory at Pulkovo of the
Russian Academy of Sciences, Saint Petersburg,
196140, Russia}
\affiliation{Institute of Physics, Nanotechnology and
Telecommunications, Peter the Great Saint Petersburg
Polytechnic University, Saint Petersburg, 195251, Russia}

\author{
V.~M.~Mostepanenko}
\affiliation{Central Astronomical Observatory at Pulkovo of the
Russian Academy of Sciences, Saint Petersburg,
196140, Russia}
\affiliation{Institute of Physics, Nanotechnology and
Telecommunications, Peter the Great Saint Petersburg
Polytechnic University, Saint Petersburg, 195251, Russia}
\affiliation{Kazan Federal University, Kazan, 420008, Russia}

\author{
 U.~Mohideen\footnote{Umar.Mohideen@ucr.edu}}
\affiliation{Department of Physics and Astronomy, University of California, Riverside, California 92521, USA}

\begin{abstract}
We present the results of an experiment on measuring the gradient of
the Casimir force between an Au-coated hollow glass microsphere and
graphene-coated fused silica plate by means of a modified atomic force
microscope cantilever based technique operated in the dynamic regime.
These measurements were performed in high vacuum at room temperature.
The energy gap and the concentration of impurities in the graphene
sample used have been measured utilizing scanning tunnelling
spectroscopy and Raman spectroscopy, respectively. The measurement
results for the gradients of the Casimir force are found to be in
a very good agreement with theory using the polarization tensor of
graphene at nonzero temperature depending on the energy gap and
chemical potential with no fitting parameters. The theoretical
predictions of the same theory at zero temperature are experimentally
excluded over the measurement region from 250 to 517 nm. We have
also investigated a dependence of the thermal correction to the
Casimir force gradient on the values of the energy gap, chemical
potential, and on the presence of a substrate supporting the graphene
sheet. It is shown that the observed thermal effect is consistent in
size with that arising for pristine graphene sheets if the
impact of real conditions such as nonzero values of the energy gap,
chemical potential, and the presence of a substrate is included.
Implications of the obtained results to the resolution of the
long-standing problems in Casimir physics are discussed.
In addition to the paper published previously [M. Liu {\it et al}.,
Phys. Rev. Lett. {\bf 126}, 206802 (2021)], we present
measurement results for the energy gap of the graphene sample, double
the experimental data for the Casimir force, and perform a more
complete theoretical analysis.
\end{abstract}

\maketitle

\section{Introduction}

An investigation of different effects in graphene brought to light that
this material possesses a variety of unusual properties which are
of much interest to fundamental physics. It is well known that
graphene is characterized by a minimum electrical conductivity
and low absorbance expressed in terms of fundamental constants
\cite{1,2,3,4} and provides new possibilities for
experimentally testing the Klein paradox \cite{5}, the effect of
Schwinger
pair creation from vacuum in external electric field \cite{6,7,8,9}, and
the relativistic quantum Hall effect \cite{10}. This is a
consequence of graphene being a two-dimensional material which at low
energies is well described not by the Schr\"{o}dinger equation
but by the relativistic Dirac equation where the speed of light $c$
is replaced by the much lower Fermi velocity $v_F$ \cite{11,12,13}.

One of the challenging problems is the experimental and theoretical
investigation of the thermal Casimir interaction in graphene systems.
The Casimir force \cite{14} is the relativistic generalization of a
more familiar van der Waals force. This is an entirely quantum
phenomenon which originates from the zero-point and thermal
fluctuations of the electromagnetic field whose spectrum is
altered by the presence of material boundaries, no matter be they
three- or two-dimensional. The fundamental unified theory of the
van der Waals and Casimir forces was created by Lifshitz \cite{15,16}.
In the framework of this theory, the Casimir free energy and force
are expressed via the reflection coefficients of electromagnetic
fluctuations on the boundary surfaces. In the original formulation,
only the plane boundaries were considered but currently the
Lifshitz theory is generalized to the case of arbitrarily shaped
bodies \cite{16a,16b,16c,17,18}.

Precise measurements of the thermal Casimir force between metallic
test bodies using the present-day laboratory techniques revealed
a puzzling problem. In many experiments performed by different
experimental groups it was found that the predictions of the
Lifshitz theory come into conflict with the measurement data if
the much-studied relaxation properties of conduction electrons
at low frequencies are taken into account in computations
\cite{19,20,21,22,23,24,25,26,27,28,29,30,31} (see also
monograph \cite{32} and reviews \cite{33,34,35}).
Note that in Ref.~\cite{35a} an agreement was obtained by subtracting a hypothetical
electrostatic force between a centimeter-size spherical lens and a plate
which was 10 times larger than the Casimir force. This result, however,
ignored imperfections of the lens surface which have an important effect
on the measured force \cite{35b}.
What is even
more surprising, an agreement between experiment and theory is
restored if computations are performed with simply discarded
relaxation properties of conduction electrons
\cite{19,20,21,22,23,24,25,26,27,28,29,30,31,32,33,34,35}.
Specifically, experiments using magnetic metal surfaces \cite{24,25,26}
and isoelectronic difference force measurements \cite{27} have reconfirmed
this conclusion with an extraordinary precision.

It should be  emphasized that the reflection coefficients
used in the standard Lifshitz theory are expressed via the
dielectric permittivities of boundary materials which in turn are
found from the available optical data for the complex index of
refraction \cite{36} extrapolated down to zero frequency.
 The best known method for extrapolation is by means of the Drude
model. Under certain assumptions, this model can be derived from
Boltzmann transport theory or the Kubo formula and finds full
verification in the area of electromagnetic phenomena other than
the Casimir effect \cite{37}. The Drude model takes  proper account
of the relaxation properties of conduction electrons in metals
by means of the temperature-dependent relaxation parameter.
However, the Lifshitz theory using the Drude model predicts a
relatively large thermal effect in the Casimir force at short
separation distances \cite{38} which was excluded by the
experiments mentioned above.

Graphene provides great advantages for the resolution of this problem.
The point is that at energies below a few electron volts
characteristic for the Casimir force at separations exceeding 100~nm
graphene can be considered in the framework of the Dirac model as
a set of massless or very light electronic quasiparticles. The
response function of such a simple system to the electromagnetic
field can be found on the basis of the first principles of quantum
electrodynamics at nonzero temperature without resort either to
phenomenological approaches or simplified models.

There is an extensive literature on the theory of the Casimir
interaction in graphene systems using the Kubo formalism,
density-density correlation functions, two-dimensional Drude
and other models
\cite{39,40,41,42,43,44,45,46,47,48,49,50,51,52,53,54,55,56}.
Specifically, using the formalism of correlation functions in the
random phase approximation, which is ultimately equivalent to the
Lifshitz theory, G\'{o}mez-Santos predicted a large thermal effect
in the Casimir interaction between two parallel graphene sheets
even at separations of tens of nanometers at room temperature
\cite{40}. This prediction relates to an order of magnitude
shorter separations compared to the
thermal effect between metallic plates predicted using
the Drude model which was already excluded experimentally
\cite{19,20,21,22,23,24,25,26,27,28,29,30,31,32,33,34,35}.

The question arises on whether or not an unusually big thermal
effect exists for graphene. This question should be answered both
theoretically and experimentally. Rigorous theoretical description
of the Casimir interaction in graphene systems is based on the
Lifshitz theory supplemented by the response function of
graphene to quantum fluctuations. The latter is given by the
polarization tensor of graphene which can be found in the
framework of the Dirac model (see, e.g., Ref.~\cite{57}).

Real graphene sheets are characterized by some value of the
chemical potential, which depends on the concentration of
impurities, and of the energy gap which is caused by structural defects,
impurities, interelectron interactions and the presence of a
substrate \cite{12,13}. The exact polarization tensor of graphene
at zero temperature was found in Ref.~\cite{58} and at nonzero
temperature in Ref.~\cite{59} (the latter results are valid only
at the pure imaginary Matsubara frequencies). The exact
expressions for the polarization tensor of gapped graphene valid
over the entire plane of complex frequencies, including the real
frequency axis, was found in Ref.~\cite{60} and generalized
for the case of nonzero chemical potential in Ref.~\cite{61}.

The formalism of the polarization tensor was used to investigate
the thermal Casimir and Casimir-Polder forces in graphene
systems \cite{62,63,64,65,66,67,68,69,70,71,72,73,74,75}.
Specifically, in Ref. \cite{64} it was shown that the polarization
tensor leads to more exact results than several phenomenological
approaches used in the literature. According to the results of
Ref. \cite{71}, the formalisms of the polarization tensor and
of the density-density correlation functions are eventually
equivalent. In fact, from the exact components of the
polarization tensor it has been possible to find the
respective density-density correlation functions at nonzero
temperature which were not known until then. Most
importantly, calculations using the polarization tensor confirmed
\cite{64,72} the prediction of an unusually big thermal effect
in the Casimir force from graphene at short separations \cite{40}.
Thus, an experimental discovery of this interesting effect has
assumed great importance for both fields of graphene and
Casimir research.

The first experiment on measuring the Casimir force between an
Au-coated sphere and a graphene-coated SiO$_2$ film deposited
on a Si substrate was performed using an atomic force microscope
based technique
operated in the dynamic regime \cite{76}. The measurement
results were found in good agreement with theory using the
polarization tensor of graphene \cite{77}. Because of the thin
SiO$_2$ film used, it was not possible, however, to separate
the unusual thermal effect from the total force gradient.
According to Ref. \cite{69}, observation of the thermal effect
from graphene would become possible by increasing the thickness
of an underlying SiO$_2$ film.

Using this approach, the thermal Casimir interaction from graphene
was recently measured in the configuration of an Au-coated sphere
and a graphene sheet deposited on thick SiO$_2$ substrate \cite{78}.
The measured gradients of the thermal Casimir force were found to
be in a very good agreement with theoretical predictions calculated
using the polarization tensor accounting for the chemical
potential of graphene determined by means of Raman spectroscopy.
An estimated range of the energy gap values was included
as a part of the theoretical error. By comparing with respective
theoretical results at zero temperature, an unusual thermal
effect from graphene was reliably demonstrated over the separation
region between a sphere and a graphene sheet from 250 to 590 nm
at room temperature.

In this paper, we present additional experimental information and
a more complete theoretical analysis regarding the experiment
on measuring the thermal Casimir interaction from graphene. While
the conclusions made in Ref.~\cite{78} were based on one
measurement set consisting of 21 runs with a step in separation
distances of 1 nm, we have now performed the second measurement
set and made an averaging procedure over a more representative
wealth of evidence which includes 42 runs. Another important
innovation is that the value of the energy gap for a graphene
sample used in the experiment was measured by means of scanning
tunneling spectroscopy. As a result, it has been possible
to compute the theoretical force gradients using the polarization
tensor with the definite values of both the chemical potential
and the energy gap of graphene rather than include an estimated
range of the energy gap values in the theoretical error as was
done in Ref.~\cite{78}. Although the measured value of the energy
gap turned out to be somewhat outside the range estimated in
Ref.~\cite{78}, we have clearly confirmed the presence of an
unusual thermal effect in the graphene sample used within the
separation region from 250 to 517~nm.

On the theoretical side, we have performed calculations
elucidating the physical nature of the unusually big thermal
effect in the Casimir interaction from graphene at short
separations and its dependence on the chemical potential,
energy gap and the presence of a substrate for real graphene
samples. The case of a pristine graphene was also considered.
A comparison between experiment and theory was made on the basis
of first principles of quantum electrodynamics at nonzero
temperature with no fitting parameters and a very good agreement
was demonstrated. Implications of the obtained results to a
long-standing problem of the thermal Casimir force between
metallic test bodies are discussed.

The paper is organized as follows. In Sec.~II, we consider the
experimental procedures used for measuring the gradient of the
Casimir force between an Au-coated sphere and a graphene-coated
SiO$_2$ substrate. Section III describes measurements of the
impurity concentration and energy gap in the experimental graphene
sample. In Sec.~IV, theory of the Casimir interaction using the
polarization tensor of graphene is briefly considered in application
to the experimental configuration. In Sec.~V, we calculate the magnitude
of the unusually big thermal effect in different graphene systems
and elucidate its physical nature. Section VI contains the
comparison between experiment and theory. In Sec.~VII, the reader
will find our conclusions and a discussion of the obtained
results and their implications.

%%%%%%%%%%%%%%%%%%%%%%%%%%%%%%%%%
\section{Measuring the Casimir force gradient from graphene
{\protect \\} using a
custom  atomic force microscope cantilever based setup in the dynamic regime}

Measurements of the gradient of the Casimir force between an Au-coated hollow
glass microsphere and a graphene-coated fused silica glass (SiO$_2$) plate have
been performed by means of a custom built atomic force microscope (AFM)
cantilever based
technique operated in the dynamic regime at a temperature $T=294.0\pm 0.5~$K in high
vacuum below $9\times 10^{-9}~$Torr. Similar setups have already been used
in previous experiments on measuring the gradient of the Casimir force between
metallic surfaces \cite{23,24,25,26,28,29,30} (the schematic can be found in
Fig.~1 of Ref.~\cite{30} but here the UV- and Ar-ion cleaning  is not used
to avoid damaging the graphene sheet). Below we consider only
the most important novel features connected with the use of graphene sample.

The main test body in this experiment was made from a large-area graphene
sheet which was chemical vapor deposition grown on a Cu foil \cite{79}.
This sheet was transferred onto a polished JGS2 grade fused silica double side
optically polished substrate of 10~cm diameter and 0.05~cm thickness \cite{80}.
This was made through an electrochemical delamination procedure \cite{79,81}.
Then a $1\times 1~\mbox{cm}^2$ piece of the graphene-coated fused silica
substrate was cut from the entire sample and used as the test body in measuring
the force gradient. After the force gradient measurements have been performed,
the rms roughness of the graphene sheet on a fused silica substrate was
measured to be $\delta_g=1.5\pm 0.1~$nm by means of an AFM. This is used in
Sec.~VI for comparison between experiment and theory.

The second test body is an Au-coated hollow glass microsphere with the
diameter $2R=120.7\pm 0.1~\mu$m measured by means of a scanning electron
microscope. In doing so the thickness of Au coating was measured to be
$120 \pm 3~$nm using an AFM. After the experi\-ment was completed, the rms
roughness of the Au coating on the sphere $\delta_s=0.9\pm 0.1~$nm was
measured by means of an AFM.

A hollow glass microsphere is attached to the end of an Au-coated tipless AFM
cantilever using silver epoxy and then coated with Au \cite{82}.
Before attaching the sphere
and Au coating, the cantilever spring constant was reduced through chemical etching
(see Ref.~\cite{30} for details). As a result, the corresponding resonant
frequency of the cantilever was decreased from $5.7579\times 10^4$ to
$3.5286\times 10^4~$rad/s by etching in 60\% potassium hydroxide solution at
$75^{\circ}$C with stirring for 100~s. Note also that prior to etching the
cantilever was washed in a buffered oxide etch solution and deionized water
for 1~min each. After the Au coating, the resonant frequency of the complete
cantilever-sphere system in vacuum was measured to be
$\omega_0=6.1581\times 10^3~$rad/s.

The vacuum chamber containing the cantilever-sphere system and graphene sample
on a fused silica substrate was pumped using an oil free scroll pump and
then followed by a turbo
pump connected in series, and finally an ion pump for further pressure reduction.
During the force measurements, only the ion pump was used thereby reducing
the mechanical vibrations (see Refs.~\cite{23,28,29,30} for details).
In the dynamic measurement regime
used, the cantilever with the attached sphere was set to oscillate
 above the graphene plane. The oscillation frequency of the cantilever
and movement of the graphene sample were monitored by two fiber interferometers
with laser light sources of 1550 and 500.1~nm wavelength,
respectively. Small changes in the separation distance between sphere and
graphene due to mechanical drift during the measurement were
monitored and corrected as described in Refs.~\cite{23,29,30}.
The frequency shifts of the cantilever oscillation induced by any external force
(electric or Casimir) were recorded using a phase lock loop (PLL)
\cite{23}. In order to stay in the linear regime, the oscillation amplitude of
the cantilever was maintained at 10~nm, and the resolution of the PLL was
measured to be 55.3~mrad/s.

The total force acting on the sphere is given by
\begin{equation}
F_{\rm tot}(a,T)= F_{\rm el}(a)+F(a,T).
\label{eq1}
\end{equation}
\noindent
Here, $F_{\rm el}$ is the electric force caused by the constant voltages $V_i$
applied to the graphene sheet using ohmic contacts while the sphere remains
grounded and by the residual potential difference $V_0$, $F(a,T)$ is the
Casimir force, $a$ is the separation distance between the sphere and graphene
sheet, and $T$ is the temperature.

Under the influence of an external force (\ref{eq1}) the resonant frequency
$\omega_0$ of the
cantilever-sphere system is modified to $\omega_r(a,T)$ and the frequency
shift
\begin{equation}
\Delta\omega(a,T)=\omega_r(a,T)-\omega_0
\label{eq2}
\end{equation}
\noindent
was recorded by the PLL at every 0.14~nm while the graphene-coated fused silica
plate was moved by the piezo actuator toward the sphere starting at the maximum
separation.  Using interpolation, the values of $\Delta\omega$
can be recalculated with a step of 1~nm.
We recall that all measurements were performed at constant
$T=294~$K. The argument $T$ in the Casimir force is discussed during comparison
with the theory in Secs.~IV--VI.

In the linear regime the frequency shift (\ref{eq2}) is given by \cite{23,83}
\begin{equation}
\Delta\omega(a,T)=-CF_{\rm tot}^{\prime}(a,T)=
-CF_{\rm el}^{\prime}(a)-CF^{\prime}(a,T).
\label{eq3}
\end{equation}
\noindent
Here, the calibration constant $C=\omega_0/(2k)$, $k$ is the spring constant of
the cantilever, and the gradient of the electrostatic force in a sphere-plate
geometry is given by \cite{23,32}
\begin{eqnarray}
&&
F_{\rm el}^{\prime}(a)=X^{\prime}(a,R)(V_i-V_0)^2,
\label{eq4} \\
&&
X^{\prime}(a,R)=\frac{2\pi\epsilon_0}{\sqrt{a(2R+a)}}
\sum_{n=1}^{\infty}{\rm csch}(n\tau)\left\{n\coth(n\tau)
\right.
\nonumber \\
&&~~\left.
\times[n\coth(n\tau)-\coth\tau]-{\rm csch}^2\tau+
n^2{\rm csch}^2(n\tau)
\right\},
\nonumber
\end{eqnarray}
where $\cos\tau\equiv 1+a/R$, $\epsilon_0$ is the permittivity of vacuum,
and the absolute separations between the zero levels of roughness on the
sphere and graphene surfaces are determined from
\begin{equation}
a=z_{\rm piezo}+z_0,
\label{eq5}
\end{equation}
\noindent
where $z_{\rm piezo}$ is the distance moved by the  graphene-coated plate
and $z_0$ is the closest separation between this plate and the sphere.

As a result, the gradients of the Casimir force can be expressed using Eq.~(\ref{eq3})
via the measured frequency shift
\begin{equation}
F^{\prime}(a,T)=-\frac{1}{C}\Delta\omega(a,T)-F_{\rm el}^{\prime}(a),
\label{eq6}
\end{equation}
\noindent
where all the necessary parameters, $C$, $z_0$, and $V_0$ are determined by means
of electrostatic calibration which is performed simultaneously with measurements
of the frequency shifts (see Refs.~\cite{23,32} for details).

%%%%%%%%%%%%%%%%%%%%%%__Fig._1__%%%%%%%%%%%%%%%%%%
\begin{figure}[!b]
\vspace*{-6.7cm}
\centerline{\hspace*{0.6cm}
\includegraphics[width=5.0in]{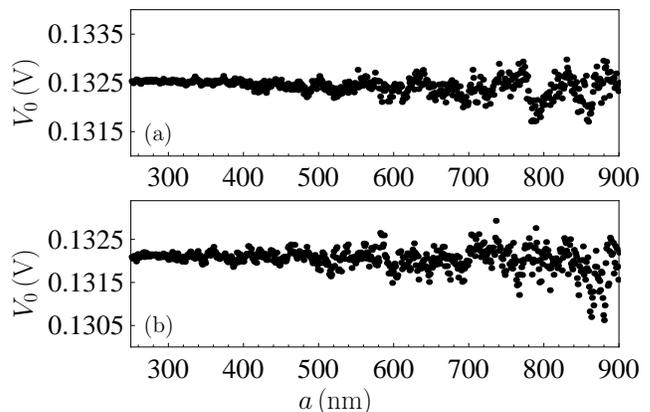}}
\vspace*{-6.5cm}
\caption{\label{fg1} The residual potential difference between an Au-coated
sphere and a graphene-coated fused silica plate is shown by the
dots as a function of separation (a) for the first and (b) for the
second measurement sets. }
\end{figure}
%%%%%%%%%%%%%%%%%%%%%%%%%%%%%%%%%%%%%%%%%%%%%%%%%%%
For this purpose, in the first measurement set reported in Ref.~\cite{78}, ten
different voltages from 0.083~V to 0.183~V with a step 0.01~V but with exception
of 0.133~V and 11 voltages equal to 0.133~V were applied to the graphene
sheet. At each separation $a$ between the graphene-coated plate and the sphere
21 values of the frequency shift $\Delta\omega$ were measured and the value
of $V$ giving
the maximum in the parabolic dependence of $\Delta\omega$ on $V_i$ in
Eqs.~(\ref{eq3}) and (\ref{eq4}), determining the value of $V_0$, was found
with the help of a $\chi^2$-fitting procedure. Using the same fit, from the
curvature  of the parabola mentioned above we have determined the values of $z_0$ and $C$.
In Fig.~\ref{fg1}(a) taken from Ref.~\cite{78}
the obtained values of $V_0$ are shown as the function of
separation between a sphere and a graphene-coated plate
for the first measurement set.
To check that the obtained values of $V_0$ do not depend on separation, we
have performed the best fit of $V_0$ to the straight line $V_0=d+\theta a$,
where $a$ is measured in nanometers, and found that
$d^{(1)}=0.1326~$V and $\theta^{(1)}=-2.73\times 10^{-7}~$V/nm \cite{78}.
This demonstrates an independence of $V_0$ on $a$ in this set of measurements
up to a high precision. The mean value of $V_0$ in the first measurement set
was $V_0^{(1)}=0.1324~$V.

In a similar way, the values of $z_0$ and $C$ were determined from the fitting
procedure at each separation and found to be separation-independent leading
to the mean values $z_0^{(1)}=236.9\pm 0.6~$nm and
$C^{(1)}=(4.599\pm 0.003)\times 10^5~$s/kg.

As mentioned above, at each separation the frequency shift $\Delta\omega$
was measured for 21 times with different applied voltages. The respective
experimental values of the gradient of the Casimir force were calculated by
Eq.~(\ref{eq6}) and their mean values $F_{(1)}^{\prime}(a,T)$ were found with
a step of 1~nm. The random errors of these mean values were determined at the
67\% confidence level. The systematic errors, which are mostly caused
by the errors in measuring the frequency shift indicated above, were combined
in quadrature with the random errors resulting in the total experimental
errors of the first measurement set $\Delta_{\rm expt}^{\!(1)}F^{\prime}(a,T)$.
The error in measuring the absolute separations was found to be
$\Delta a=0.6~$nm.

In addition to the first measurement set reported in Ref.~\cite{78}, second
set of measurements was performed with the same graphene sample, applied
voltages and using the same experimental procedures. This resulted in values
of the residual potential difference shown in Fig.~\ref{fg1}(b) as a function of
separation. The best fit of these values to straight line results in
$d^{(2)}=0.1322~$V and $\theta^{(2)}=-3.41\times 10^{-7}~$V/nm.
The obtained parameters are only slightly different from those
in the first set of measurements and again demonstrate an independence of the
residual potential difference on separation.
In the second measurement set the mean value of $V_0$ was determined
to be $V_0^{(2)}=0.1320~$V.

The values of the separation on contact and the calibration constant in the
second measurement set were also found to be independent of separation resulting
in the following  mean values: $z_0^{(2)}=238.8\pm 0.5~$nm and
$C^{(2)}=(4.712\pm 0.003)\times 10^5~$s/kg.

The independence of the residual potential difference
on the sphere-graphene separation in both
measurement sets confirms that in this experiment performed
in high vacuum the role of patch potentials
on an Au-coated sphere and of spurious electrostatic interactions induced
by charges on the SiO$_2$ substrate supporting graphene is negligibly small
for the separations reported here.
Similar situation holds for the experiments
\cite{19,20,21,22,23,24,25,26,27,28,29,30,31}
performed in high vacuum with two Au or Ni-coated test bodies where a smallness of the
electrostatic effects was confirmed by independent measurements employing
Kelvin-probe microscopy \cite{86a}.
Note that graphene sheet is connected to a power supply which is a reservoir for
compensating charges.
As the graphene sheet is a two-dimensional
conducting layer with high conductivity determined by the very light Dirac quasiparticles,
which is connected to a power supply,
it effectively screens out the role of possible charges on the SiO$_2$ substrate.
This is confirmed by the measurements of mean impurity concentration in graphene
presented in Sec.~III.  By contrast, the cases, where the role of patch effects can be
relatively large, are considered in the experiments of Ref.~\cite{86b} performed in
ambient air with 30\% relative humidity.

These results were used to find the experimental values of the gradients of the
Casimir force at each separation and
their mean $F_{(2)}^{\prime}(a,T)$  with a step of 1~nm.
Following the same procedure as described above, the total experimental
errors in the second measurement set $\Delta_{\rm expt}^{\!(2)}F^{\prime}(a,T)$
were determined.

In each measurement set, the total experimental error is mostly determined by the
systematic error which is almost the same for both sets. In doing so an advantage
of using the two sets of measurements is in the decreased impact of possible
accidental systematic deviations.

Finally, we have calculated the experimental gradients of the Casimir force,
$F_{\rm expt}^{\prime}(a,T)$, by averaging the mean values obtained in two
measurement sets. In a similar way, the total experimental error of the measured
gradients,   $\Delta_{\rm expt}F^{\prime}(a,T)$, was obtained by averaging the
total experimental errors found in the first and second measurement sets.

\begin{widetext}
%%%%%%%%%%%%%%%%%%%%%%__Fig._2__%%%%%%%%%%%%%%%%%%
\begin{figure*}[!t]
\vspace*{-4.7cm}
\centerline{%\hspace*{6.5cm}
\includegraphics[width=8.20in]{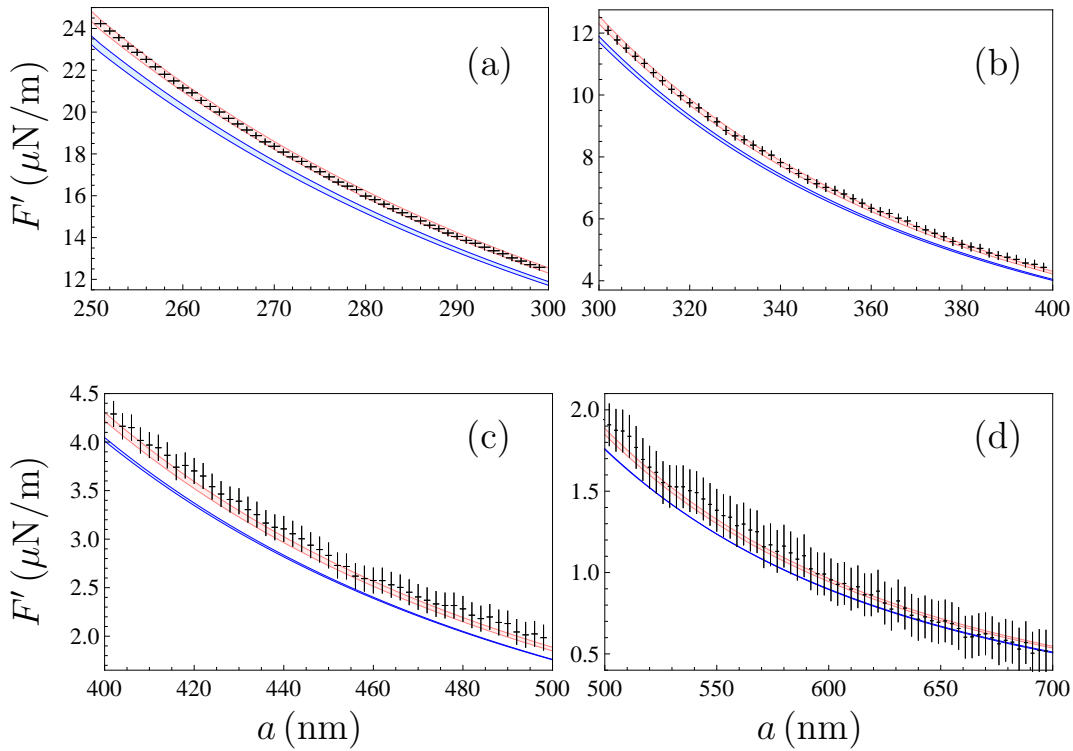}}
\vspace*{-13.8cm}
\caption{\label{fg2} The mean gradient of the Casimir force obtained from the
two measurement sets is shown by the crosses as a function of
separation within four separation intervals. The upper and lower
theoretical bands are computed at the laboratory temperature
$T=294~$K and at $T=0~$K, respectively (see the text for further
discussion). }
\end{figure*}
%%%%%%%%%%%%%%%%%%%%%%%%%%%%%%%%%%%%%%%%%%%%%%%%%%%
\end{widetext}

The measurement results for  $F_{\rm expt}^{\prime}(a,T)$ obtained from the two
sets of measurement are shown as crosses in Figs.~\ref{fg2}(a)--\ref{fg2}(d)
over the separation range from 250 to 700~nm. The vertical and horizontal
arms of the crosses have the lengths $2\Delta_{\rm expt}F^{\prime}(a,T)$ and
$2\Delta a$, respectively, determined by the total experimental errors.
For visual clarity, we have indicated all data points in Fig.~\ref{fg2}(a),
each second data point in Figs.~\ref{fg2}(b) and \ref{fg2}(c), and each third
data point  in Fig.~\ref{fg2}(d). The top and bottom bands indicated in
Fig.~\ref{fg2} refer to the comparison between experiment and theory which is
discussed in Sec.~VI. Note that the minimum separation distance of 250~nm chosen
in the experimental data reported here
 is typical for measurements of the Casimir force by means
of an atomic force microscope in the dynamic mode \cite{23,24,25,26,28,29,30}.
This is done in order do not enter a nonlinear regime of the oscillator system
used. On the theoretical
side, the relative thermal effect in the Casimir interaction from graphene
becomes more sensible just at $a>250~$nm (see below in Sec.~V, Fig.~6).
%%%%%%%%%%%%%%%%%%%%%__Fig._2___%%%%%%%%%%%%%%%%%%%%%%%%%%%

%%%%%%%%%%%%%%%%%%%%%%%%%%%%%%%%%
\section{Measurements of the impurity concentration and energy gap of graphene
sheet deposited on a fused silica substrate}

It is well known that real graphene sheets are characterized
by some small but nonzero mass of electronic quasiparticles which
leads to the presence of an energy gap $\Delta$ in their spectrum
\cite{12,13}. In a similar manner, real graphene samples contain
some fraction of impurities and, as a result, are characterized
by some nonzero chemical potential $\mu$ \cite{12,13}. For a
pristine graphene sheets it holds $\Delta=\mu=0$. One should know
the values of both the energy gap and chemical potential in order
to calculate the response of a graphene sample to electromagnetic
fluctuations. Because of this, it is desirable to determine both
of them before comparing measurement results of the Casimir
interaction in graphene systems with theoretical predictions.

We begin with determining the value of the chemical potential
which is
caused by the concentration of impurities. The impurity
concentration in the graphene sample was determined using Raman
spectroscopy after the measurement of the Casimir force gradient.
The Raman spectroscopy was carried out using a Horiba Labram HR 800
system with 532 nm laser excitation (Laser Quantum, 65~mW power).
A 100x objective lens with NA = 0.9, which renders a laser spot size
of ~0.4 $\mu$m$^2$ and corresponding spot diameter of 709 nm was used.
The measurements were done at a temperature of 294 $\pm$ 0.5 K. The
spectrometer used a grating with 600 lines/mm to ensure the spectral
range from  1450~cm$^{-1}$ to 2900~cm$^{-1}$ which includes both G and
2D peaks of graphene. The spectral resolution was maintained at
2~cm$^{-1}$ for the precise detection of the G peak blue shift. The
spectra were collected at 18 equidistant points on the sample in order
to understand the spatial distribution of the impurity concentration.
Prior to acquiring the spectra, to ensure that the sample was
positioned at the focal plane the signal intensity was maximized by
adjusting the focus of the microscope. The spectra collected at each
point are the accumulated results over 10 acquisitions with each
acquisition spanning over 10~s.

\begin{figure}[!b]
\vspace*{-5.5cm}
\centerline{\hspace*{0.3cm}
\includegraphics[width=5.50in]{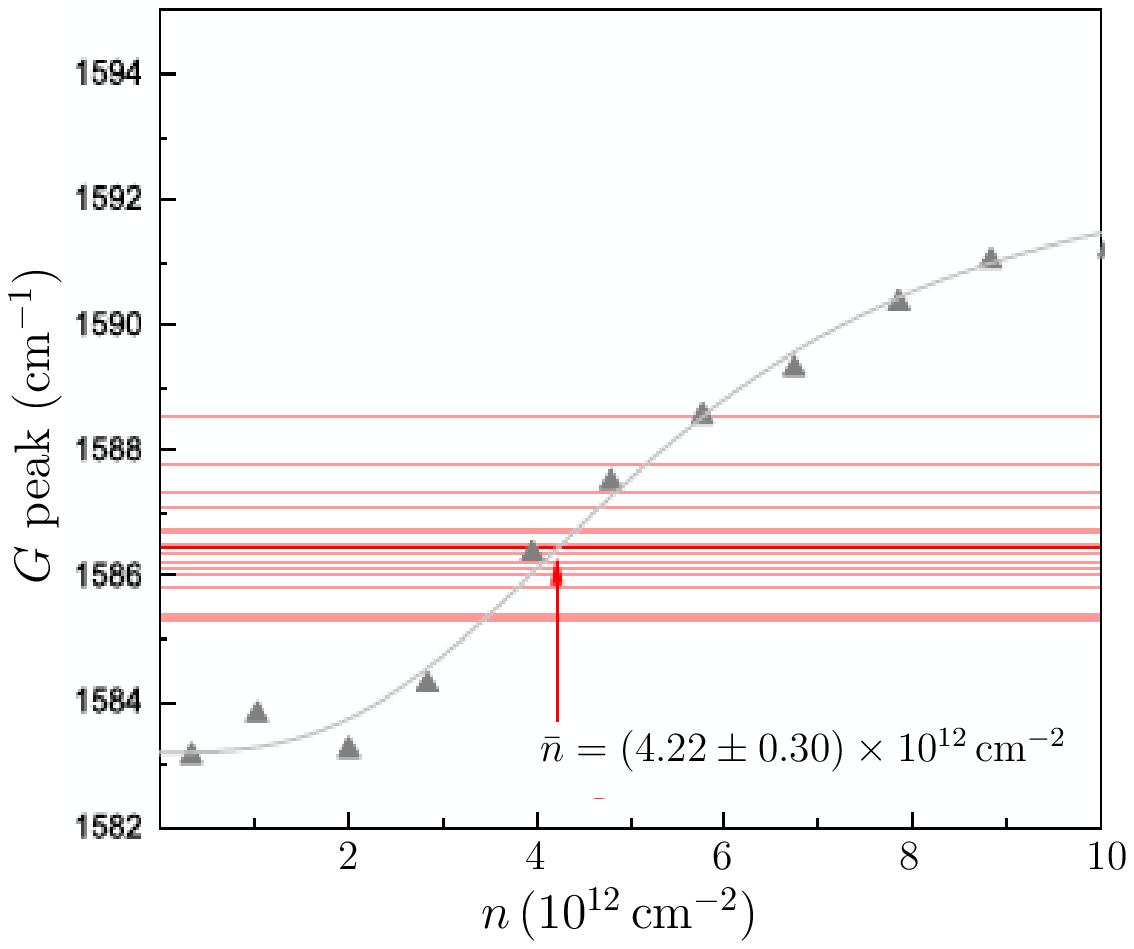}}
\vspace*{-8.cm}
\caption{\label{fg3}  The measured $G$-peak values from Raman spectroscopy
compared to the charge concentration in graphene from
Ref.~\cite{84}. The solid triangles are the values from
Ref.~\cite{84} and the solid gray line is a best fit to the data.
The $G$-peak values measured at equidistant points on the sample
are shown by horizontal lines. The intersection identifies the
impurity concentration. The average impurity concentration is
shown by the arrow. }
\end{figure}

\begin{figure}[!t]
\vspace*{-2.7cm}
\centerline{\hspace*{0.3cm}
\includegraphics[width=10.50in]{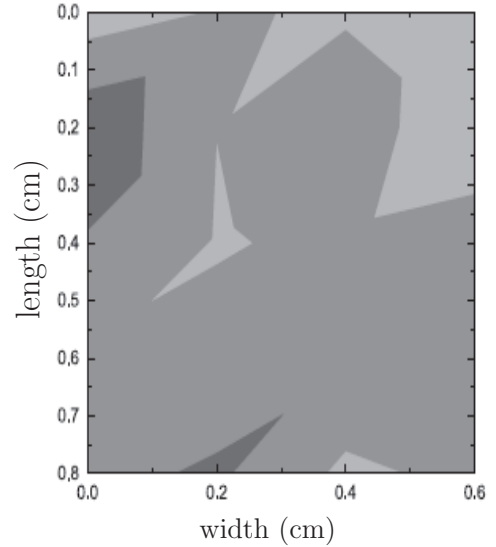}}
\vspace*{-27.cm}
\caption{\label{fg4} The impurity concentration
measured at different points over the $0.6 \times 0.8~\mbox{cm}^2$ area of
the sample from Fig.~\ref{fg3} is shown by
the light gray, gray, and dark gray regions where it varies
from $3\times 10^{12}$ to $4\times 10^{12}~\mbox{cm}^{-2}$,
$4\times 10^{12}$ to $5\times 10^{12}~\mbox{cm}^{-2}$, and
from $5\times 10^{12}$ to $6\times 10^{12}~\mbox{cm}^{-2}$, respectively.
}
\end{figure}

The $G$-peak spectra were fitted to a Lorenzian to identify
the precise location of the peak. The values of the
$G$-peak were compared to $G$-peak shifts modified by charge
concentration that are reported in Ref.~\cite{84} and the
corresponding impurity concentration was identified as shown in
Fig.~\ref{fg3}. Here, the solid triangles are the data from Ref.~\cite{84}
and the gray solid line is a fit to the data. The measured G peaks
are shown by the horizontal lines. The mean value of the impurity
concentration from all the measured G peaks is
$\bar{n}$ = ($4.2 \pm 0.3) \times 10^{12}~\mbox{cm}^{-2}$, where the random
and systematic errors were summed to obtain the maximum possible
error. Na is expected to be the dominant impurity type based on the
transfer process used \cite{81}. Figure \ref{fg4} presents a spatial
distribution of the impurity concentration from Fig.~\ref{fg3} measured over
the area $0.6 \times 0.8~\mbox{cm}^2$ of the sample.
In accordance with Fig.~\ref{fg3}, the light gray, gray, and dark gray areas
in the field of Fig.~\ref{fg4} correspond to the impurity concentrations varying
from $3\times 10^{12}$ to $4\times 10^{12}~\mbox{cm}^{-2}$,
$4\times 10^{12}$ to $5\times 10^{12}~\mbox{cm}^{-2}$, and
from $5\times 10^{12}$ to $6\times 10^{12}~\mbox{cm}^{-2}$, respectively.
%%%%%%%%%%%%%%%%%%%%%%__Fig._3__%%%%%%%%%%%%%%%%%%
%%%%%%%%%%%%%%%%%%%%%%%%%%%%%%%%%%%%%%%%%%%%%%%%%%%
%%%%%%%%%%%%%%%%%%%%%%__Fig._4__%%%%%%%%%%%%%%%%%%
%%%%%%%%%%%%%%%%%%%%%%%%%%%%%%%%%%%%%%%%%%%%%%%%%%%

The respective value of the chemical potential of our graphene
sample at zero temperature is given by \cite{85a}
\begin{equation}
\mu = \hbar v_F \sqrt{\pi\bar{n}}=0.24\pm 0.01~\mbox{eV},
\label{eq6a}
\end{equation}
\noindent
where the Fermi velocity $v_F \approx c/300$. According to
Ref.~\cite{85b}, due to the relatively large value of the chemical
potential, as in Eq.~(\ref{eq6a}), it is almost independent on
temperature. Because of this, one can use the obtained value
(\ref{eq6a}) both at zero and room temperatures.

We proceed now to measurements of the energy gap $\Delta$.
The energy gap of the graphene mounted on the fused silica was
determined using Scanning Tunneling Sprectroscopy (STS) \cite{86}.
The STS measurements were performed using a Nanosurf Nano STM. The
probe was fabricated by mechanically cutting a Pt-Ir wire,
generating a sharp tip. The graphene sample on the fused silica
substrate was cut into $5 \times 5~\mbox{mm}^2$ pieces and mounted onto
metal puck holders using a conductive adhesive. Conduction between
graphene surfaces and the metal puck was achieved by using the same
conductive epoxy.

All experiments were performed in air at $22 \pm 1^{\circ}$C.
The STM was kept in an enclosed environment on a floating optical
table to minimize thermal and vibrational fluctuations. To select
an appropriate region free from surface wrinkles and corrugations,
rough microscopic scans ($50\times 50~\mbox{nm}^2$ to
$10\times 10~\mbox{nm}^2$) of the
surface topography were performed prior to spectroscopic measurements.
The microscopic scans were performed with a bias voltage of 50~mV
and a tunneling current of 1~nA.

The spectroscopic scans were performed in ``current-voltage mode"
where the measurement position and tip-sample separation distance
were checked to be constant by monitoring the initial tunneling
current prior to the spectroscopic scans. For the spectroscopic
scans the bias voltage was linearly ramped from --1.2~V to 1.2~V
over a 100 ms period. As the experiments were carried out at room
temperature, experiments at different periods of 10~ms and 50~ms
were also tried and verified to lead to similar results. The final
spectroscopic scans were all carried out with 100~ms period and the
tunneling currents were recorded at 256 equal time intervals during
each ramp from --1.2~V to +1.2~V.  The experiment was repeated 3--4
times till a reproducible spectrogram was obtained. The entire
experiment  was repeated at 50 different positions on 3 different
samples.

\begin{figure}[!b]
\vspace*{-4.7cm}
\centerline{\hspace*{0.cm}
\includegraphics[width=4.50in]{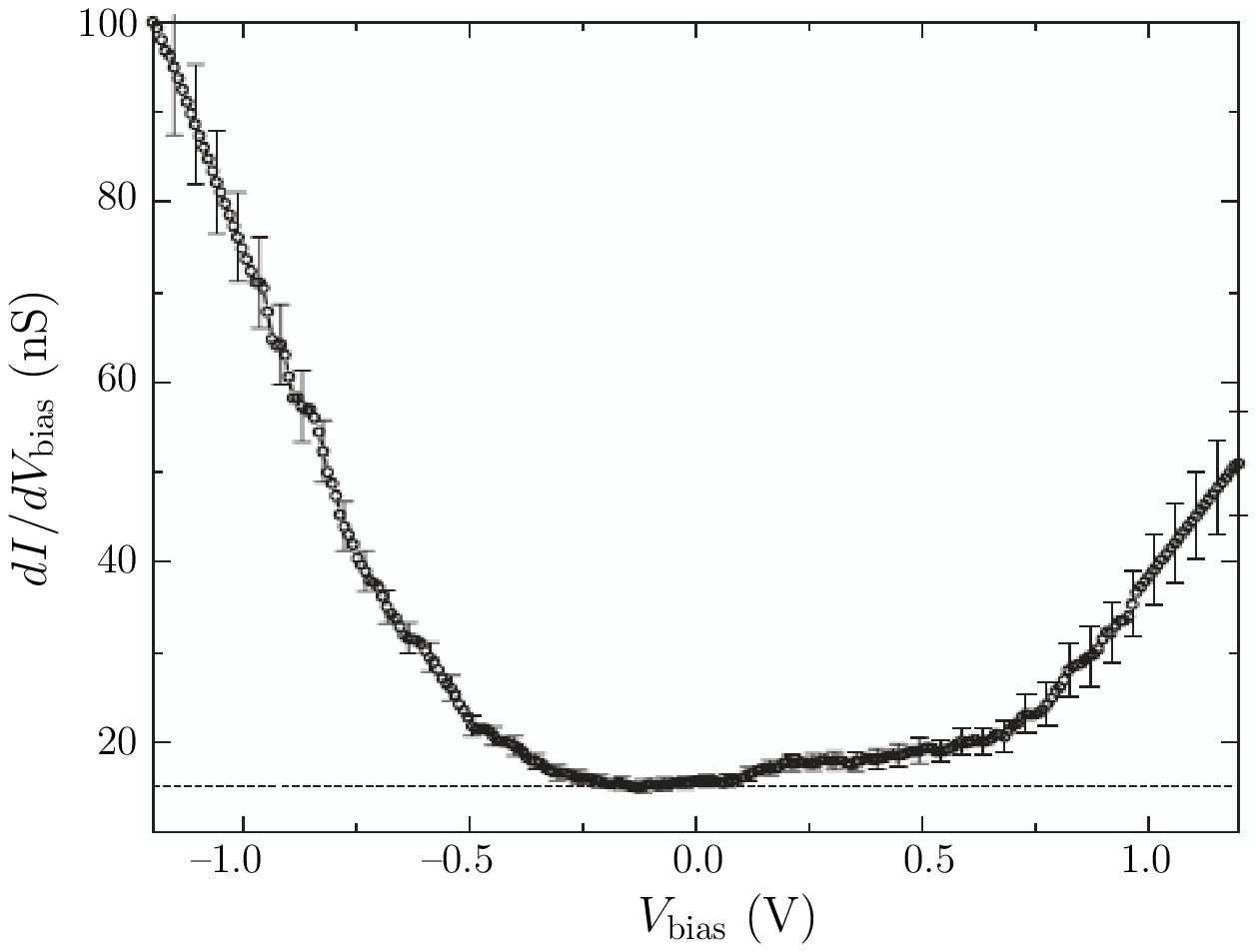}}
\vspace*{-6.8cm}
\caption{\label{fg5}  The average measured differential conductance
$dI/dV_{\rm bias}$, measured from the scanning tunneling
spectroscopy of the graphene sample as a function of the bias
voltage. For clarity the error bars are shown only at every fifth
data point. The horizontal dashed line is the minimum average
differential conductance measured. }
\end{figure}

{}From the tunneling current $I$ as a function of the bias voltage
$V_{\rm bias}$, the differential conductance, $dI/dV_{\rm bias}$ as a
function of the bias voltage was determined.  An averaged
differential conductance as a function of $V_{\rm bias}$ from the 50
different measured spectra obtained from the three different samples
is shown in Fig.~\ref{fg5}. The average minimum value of the differential
conductance measured is shown as a horizontal dashed line. A U shaped
parabolic dependence of the differential conductance with bias
voltage was observed. The V shape differential conductance with bias
voltage reported for a pristine graphene at low temperature was not
observed due to room temperature measurement as well as the presence
of an energy gap from the presence of impurities \cite{87,88} and
the mechanical strain from the substrate \cite{89,90}, both of which
modify the local density of states.
%%%%%%%%%%%%%%%%%%%%%%__Fig._5__%%%%%%%%%%%%%%%%%%
%%%%%%%%%%%%%%%%%%%%%%%%%%%%%%%%%%%%%%%%%%%%%%%%%%%

Previous Raman spectroscopy mapping of charge impurities on the
sample (see Fig.~\ref{fg4}) shows that the impurity density varies with
position leading to variations in the differential conductance at
different positions.  The substrate induced roughness also leads to
similar variations with position \cite{89,90,91}. To determine the
band edge at negative voltage bias, the differential conductance
curve in that region was extrapolated to intersect with the line of
minimum $dI/dV_{\rm bias}$ \cite{92}. The extrapolation was done using
a semilog plot of the differential conductance.
Uncertainties in the extrapolation are
recorded as uncertainties in the determination of the band edge.
This was repeated for the differential conductance curve at the
positive bias to identify the band edge in that region. The width
of the energy gap was thus determined to be
$\Delta = 0.29 \pm 0.05~$eV using this procedure.

The obtained values of the energy gap and chemical potentials are
used in Secs.~V and VI where the gradients of the Casimir force are
computed using the formalism of the polarization tensor.

%%%%%%%%%%%%%%%%%%%%%%%%%%%%%%%%%
\section{Theory of thermal Casimir interaction from graphene using the
polarization tensor in the experimental configuration}

As mentioned in Sec.~I, there are many theoretical approaches used in the
literature for a description of the Casimir interaction in graphene systems.
Here, we describe the gradient of the Casimir force in the experimental
configuration by means of the Lifshitz theory which is valid for any planar
layered structures with appropriately found reflection coefficients.
In doing so the response of Au and fused silica to the electromagnetic field
is described by their frequency-dependent dielectric permittivities,
whereas the response of graphene can be found precisely in the framework of
the Dirac model using the polarization tensor in (2+1)-dimensional space-time.
An employment of the Dirac model is fully justified because even at the
shortest separation considered in our experiment ($a=250~$nm) the characteristic
energy of the Casimir force $\hbar\omega_c=\hbar c/(2a)=0.4~$eV is well within
the application region of the Dirac model of graphene (according to recent
results it is applicable up to 3~eV \cite{93}). Because of this, one need not
 take  into consideration the absorption peak of graphene at $\lambda=270~$nm
which corresponds to much higher energy
$\hbar\omega=2\pi\hbar c/\lambda\approx 4.59~$eV.

Using the proximity force approximation (PFA) \cite{32} (corresponding error in this
experiment is very small and is taken into account below), the gradient of the
Casimir force between an Au-coated sphere and a graphene-coated SiO$_2$ plate
calculated at the laboratory temperature $T$ takes the form \cite{69,72,77}
\begin{eqnarray}
&&
F^{\prime}(a,T)=2k_BTR\sum_{l=0}^{\infty}{\vphantom{\sum}}^{\prime}
\int_0^{\infty}q_lk_{\bot}dk_{\bot}
\nonumber \\
&&~~
\times\sum_{\sigma}\left[r_{\sigma}^{-1}(i\xi_l,k_{\bot})
R_{\sigma}^{-1}(i\xi_l,k_{\bot},T)e^{2aq_l}-1\right]^{-1}\!\! .
\label{eq7}
\end{eqnarray}

In this equation, $k_B$ is the Boltzmann constant,  the prime on the
first summation sign divides the term with $l=0$ by 2,
$k_{\bot}$ is the magnitude of the wave vector projection on a graphene
plane, the Matsubara frequencies are $\xi_l=2\pi k_BTl/\hbar$,
$q_l=\sqrt{k_{\bot}^2+{\xi_l^2}/{c^2}}$, and the summation in
$\sigma$ is over the two polarizations of the electromagnetic field,
transverse magnetic ($\sigma={\rm TM}$) and
transverse  electric ($\sigma={\rm TE}$).

Now it is necessary to define the reflection coefficients $r_{\sigma}$
and $R_{\sigma}$ entering Eq.~({\ref{eq7}). In doing so, taking into account
the sufficiently thick Au-coating on the sphere and large thickness of the
SiO$_2$ plate, the sphere can be considered as all-gold and the plate ---
as a semispace \cite{32}. Then, the coefficients $r_{\sigma}$ on the
boundary between Au and vacuum take the standard form \cite{32,33,34}
\begin{eqnarray}
&&
r_{\rm TM}(i\xi_l,k_{\bot})=
\frac{\varepsilon_l^{(1)}q_l-k_l^{(1)}}{\varepsilon_l^{(1)}q_l+k_l^{(1)}},
\nonumber \\
&&
r_{\rm TE}(i\xi_l,k_{\bot})=
\frac{q_l-k_l^{(1)}}{q_l+k_l^{(1)}},
\label{eq8}
\end{eqnarray}
\noindent
where
\begin{equation}
k_l^{(n)}=k_l^{(n)}(k_{\bot})=
\sqrt{k_{\bot}^2+\varepsilon_l^{(n)}\frac{\xi_l^2}{c^2}}
\label{eq9}
\end{equation}
\noindent
and $\varepsilon_l^{(n)}=\varepsilon^{(n)}(i\xi_l)$ are the
dielectric permittivities of Au ($n=1$) and SiO$_2$ ($n=2$) calculated at the
pure imaginary Matsubara frequencies.

The reflection coefficients $R_{\sigma}$ on the
boundary between vacuum  and graphene-coated plate are more involved because the plate
material is described by the dielectric permittivity $\varepsilon _l^{(2)}$ whereas
graphene --- by the polarization tensor in (2+1)-dimensional space-time
\begin{equation}
\Pi_{\beta\gamma,l}\equiv\Pi_{\beta\gamma}(i\xi_l,k_{\bot},T,\Delta,\mu),
\label{eq10}
\end{equation}
\noindent
where $\beta,\,\gamma=0,\,1,\,2$.
This tensor depends on temperature $T$ and on the energy gap $\Delta$ and chemical
potential $\mu$ of a graphene sheet. In fact it has only two independent components
\cite{59,60,61}. It is most convenient to express the reflection coefficients via
$\Pi_{00,l}$ and  the following combination of the components
$\Pi_l$ defined as
\begin{equation}
\Pi_l=k_{\bot}^2\Pi_{\beta,\,l}^{\,\,\beta}-q_l^2\Pi_{00,l},
\label{eq11}
\end{equation}
\noindent
where $\Pi_{\beta,\,l}^{\,\,\beta}$ (the summation in $\beta$ is implied) is
the trace of the polarization tensor.

Using the above notations, the reflection coefficients $R_{\sigma}$ are given by
\cite{72,74,77}
\begin{eqnarray}
&&
R_{\rm TM}(i\xi_l,k_{\bot},T)=
\frac{\hbar k_{\bot}^2[\varepsilon_l^{(2)}q_l-k_l^{(2)}]+
q_lk_l^{(2)}\Pi_{00,l}} {\hbar k_{\bot}^2[\varepsilon_l^{(2)}q_l+k_l^{(2)}]+
q_lk_l^{(2)}\Pi_{00,l}},
\nonumber\\
&&
R_{\rm TE}(i\xi_l,k_{\bot},T)=
\frac{\hbar k_{\bot}^2[q_l-k_l^{(2)}]-
\Pi_{l}} {\hbar k_{\bot}^2[q_l+k_l^{(2)}]+\Pi_{l}}.
\label{eq12}
\end{eqnarray}

To calculate the gradient of the Casimir force using Eqs.~(\ref{eq7})--(\ref{eq12})
one needs to have the values of the dielectric permittivities $\varepsilon_l^{(n)}$
and of the quantities $\Pi_{00,l}$ and $\Pi_l$ as input information.
As mentioned in Sec.~I, the quantities $\varepsilon_l^{(n)}$ are obtained from the
measured optical data for the complex index of refraction \cite{36}.
In the case of one test body coated with a graphene sheet the reflection coefficient
$R_{\rm TE}(0,k_{\bot},T)$ turns out to be very small. Because of this, the gradients
of the Casimir force calculated by Eqs.~(\ref{eq7})--(\ref{eq12}) are almost independent
of the type of extrapolation of the optical data to zero frequency discussed in Sec.~I.
Therefore, one can safely employ the values of $\varepsilon_l^{(n)}$ available in the
literature \cite{32,33,34} obtained using any extrapolation (i.e., taking into
account or disregarding the relaxation properties of conduction electrons)
leading to coinciding
results. This gives the possibility to consider the reflection coefficients $r_{\sigma}$
as independent of $T$.

The quantities $\Pi_{00,l}$ and $\Pi_l$ are conveniently presented as the sums of
two contributions
\begin{equation}
\Pi_{00,l}=\Pi_{00,l}^{(0)}+\Pi_{00,l}^{(1)},
\quad
\Pi_{l}=\Pi_{l}^{(0)}+\Pi_{l}^{(1)}.
\label{eq13}
\end{equation}
\noindent
The first terms on the right-hand sides of these equations are related to the
polarization tensor of an undoped graphene  with $\mu=0$, arbitrary value of the
energy gap $\Delta$,  at zero temperature $T=0$, but calculated at the pure
imaginary Matsubara frequencies $\omega=i\xi_l$.
This means that the quantities $\Pi_{00,l}^{(0)}$ and $\Pi_{l}^{(0)}$ take into
account only an implicit dependence of the polarization tensor on temperature
through the Matsubara frequencies.
The second terms on the right-hand sides of Eq.~(\ref{eq13}) result from an
explicit dependence of the polarization tensor on temperature $T$ and on the
chemical potential $\mu$. In so doing they also depend on $\Delta$.

The explicit expressions for $\Pi_{00,l}^{(0)}$ and $\Pi_{l}^{(0)}$ are given by
 \cite{58,59}
\begin{equation}
\Pi_{00,l}^{(0)}=\frac{\alpha\hbar k_{\bot}^2}{\tilde{q}_l}\Psi(D_l),
\quad
\Pi_{l}^{(0)}=\alpha\hbar\tilde{q}_l\Psi(D_l),
\label{eq14}
\end{equation}
\noindent
where $\alpha=e^2/(\hbar c)$ is the fine structure constant and
\begin{eqnarray}
&&
\tilde{q}_l=\left(\frac{v_F^2}{c^2}k_{\bot}^2+\frac{\xi_l^2}{c^2}
\right)^{1/2},
\quad
D_l=\frac{\Delta}{\hbar c\tilde{q}_l},
\nonumber \\
&&
\Psi(x)=2\left[x+(1-x^2)\arctan\frac{1}{x}\right].
\label{eq15}
\end{eqnarray}

The exact expressions for $\Pi_{00,l}^{(1)}$ and $\Pi_{l}^{(1)}$ are more involved.
They can be conveniently presented in the form \cite{61,74}
\begin{eqnarray}
&&
\Pi_{00,l}^{(1)}=\frac{4\alpha\hbar c^2\tilde{q}_l}{v_F^2}
\int_{D_l}^{\infty}du \left(\sum_{\kappa=\pm 1}
\frac{1}{e^{B_lu+\kappa\frac{\mu}{k_BT}}+1}\right)
\nonumber \\
&&
~\times\left[1-{\rm Re}\frac{1-u^2+2i\gamma_lu}{\left(1-u^2
+2i\gamma_lu+D_l^2-\gamma_l^2D_l^2\right)^{1/2}}
\right],
\nonumber \\[-1.5mm]
&& \label{eq16}\\
&&
\Pi_{l}^{(1)}=-\frac{4\alpha\hbar \tilde{q}_l\xi_l^2}{v_F^2}
\int_{D_l}^{\infty}du \left(\sum_{\kappa=\pm 1}
\frac{1}{e^{B_lu+\kappa\frac{\mu}{k_BT}}+1}\right)
\nonumber \\
&&
~\times\left[1-{\rm Re}\frac{(1+i\gamma_l^{-1}u)^2+(\gamma_l^{-2}-1)
D_l^2}{\left(1-u^2
+2i\gamma_lu+D_l^2-\gamma_l^2D_l^2\right)^{1/2}}
\right],
\nonumber
\end{eqnarray}
\noindent
where $\gamma_l\equiv\xi_l/(c\tilde{q}_l)$ and
$B_l\equiv\hbar c\tilde{q}_l/(2k_BT)$.

To gain better insight into the meaning of two contributions in Eq.~(\ref{eq13}),
we note that the quantity (\ref{eq16}) can be also presented as
\begin{eqnarray}
&&
\Pi_{00,l}^{(1)}=A_{00,l}(\Delta,\mu)+B_{00,l}(T,\Delta,\mu),
\nonumber \\
&&
\Pi_{l}^{(1)}=A_{l}(\Delta,\mu)+B_{l}(T,\Delta,\mu),
\label{eq17}
\end{eqnarray}
\noindent
where $A_{00,l}$ and $A_l$ do not depend on $T$ and go to zero with vanishing
$\mu$ whereas $B_{00,l}$ and $B_l$  go to zero with vanishing $T$.

In practical computations, it is convenient to consider separately the contributions
to Eq.~(\ref{eq7}) with $l=0$ and with all $l\geqslant 1$. Thus, from Eqs.~(\ref{eq13}),
(\ref{eq14}), and (\ref{eq16}) for $l=0$ one obtains
\begin{widetext}
\begin{eqnarray}
&&
\Pi_{00,0}=\alpha\hbar c\frac{k_{\bot}}{v_F}\,
\Psi(D_0)+
\frac{8\alpha k_BTc}{v_F^2}\sum_{\kappa=\pm 1}
\ln\left(e^{\kappa\frac{\mu}{k_BT}}+
e^{-\frac{\Delta}{2k_BT}}\right)
\nonumber \\
&&
~
-\frac{4\alpha\hbar c k_{\bot}}{v_F}
\int_{D_0}^{\sqrt{1+D_0^2}}du
\left(\sum_{\kappa=\pm 1}
\frac{1}{e^{B_0u+\kappa\frac{\mu}{k_BT}}+1}\right)
\frac{1-u^2}{\sqrt{1-u^2+D_0^2}},
\nonumber \\
&&
\Pi_{0}=\alpha\hbar \frac{v_Fk_{\bot}^3}{c}\,
\Psi(D_0)
\label{eq18} \\
&&
~
+4\alpha\hbar\frac{ v_Fk_{\bot}^3}{c}
\int_{D_0}^{\sqrt{1+D_0^2}}du
\left(\sum_{\kappa=\pm 1}
\frac{1}{e^{B_0u+\kappa\frac{\mu}{k_BT}}+1}\right)
\frac{-u^2+D_0^2}{\sqrt{1-u^2+D_0^2}},
\nonumber
\end{eqnarray}
\end{widetext}
\noindent
where, according to the above notations,
$D_0=\Delta/(\hbar v_Fk_{\bot})$ and
$B_0=\hbar v_Fk_{\bot}/(2k_BT)$.

At room temperature and $a>100~$nm one can obtain also much simpler approximate
expressions for $\Pi_{00,l}$ and $\Pi_l$ with $l\geqslant 1$ than the exact ones
given by Eqs.~(\ref{eq13}), (\ref{eq14}), and (\ref{eq16}).
For this purpose the condition $\xi_1\gg v_F/(2a)$ should be used leading to
\cite{72,74}
\begin{eqnarray}
&&
\Pi_{00,l}\approx\alpha\hbar c \frac{k_{\bot}^2}{\xi_l}\,
\left[\Psi\!\left(\frac{\Delta}{\hbar\xi_l}\right)+
{Y}_l(T,\Delta,\mu)\right],
\nonumber \\
&&
\Pi_{l}\approx\alpha\hbar\xi_l \frac{k_{\bot}^2}{c}\,
\left[\Psi\!\left(\frac{\Delta}{\hbar\xi_l}\right)+
{Y}_l(T,\Delta,\mu)\right],
\label{eq19}
\end{eqnarray}
\noindent
where
\begin{eqnarray}
&&
{Y}_l(T,\Delta,\mu)=2\int_{\Delta/(\hbar\xi_l)}^{\infty} du
\left(\sum_{\kappa=\pm 1}
\frac{1}{e^{B_lu+\kappa\frac{\mu}{k_BT}}+1}\right)
\nonumber \\
&&~~~~~~~~
\times
\frac{ u^2+\left(\frac{\Delta}{\hbar\xi_l}\right)^2}{u^2+1}.
\label{eq20}
\end{eqnarray}

It was shown \cite{72,74} that
numerical computations of the Casimir force using the exact expressions
(\ref{eq13}), (\ref{eq14}), and (\ref{eq16}) for the polarization tensor and,
alternatively, the exact expressions (\ref{eq18}) for $l=0$ but the
approximate expressions (\ref{eq19}) for $l\geq 1$ at room temperature and
$a\geq 100\,$nm lead to computational results
differing by less than 0.01\%.

Below we consider not only the gradient of the Casimir force
between an Au-coated sphere and graphene-coated substrate but also
the thermal correction to it defined as
\begin{equation}
\Delta_T F^{\prime}(a,T)=F^{\prime}(a,T)-F^{\prime}(a,0).
\label{eq21}
\end{equation}
\noindent
The gradient of the Casimir force at zero temperature, $F^{\prime}(a,0)$,
can be calculated by the
Lifshitz formula (\ref{eq7}) where a summation over the discrete Matsubara
frequencies should be replaced with {an} integration over the axis
of pure imaginary frequency according to
\begin{equation}
k_BT\sum_{l=0}^{\infty}{\vphantom{\sum}}^{\prime}\to
\frac{\hbar}{2\pi}\int_{0}^{\infty}d\xi.
\label{eq22}
\end{equation}
\noindent
This means that in Eq.~(\ref{eq7}) one should replace
 $\xi_l$ and $q_l$  with $\xi$ and $q$.
 The respective replacements, which also include the changes of
 $k_l^{(n)}$ for $k^{(n)}$, $\tilde{q}_l$ for $\tilde{q}$,
 $\Pi_{00,l}$ for $\Pi_{00}$, and $\Pi_l$ for $\Pi$, should be made
in Eqs.~(\ref{eq8})--(\ref{eq16}).

To calculate the gradients of the Casimir force at zero temperature,
one also needs explicit expressions for the quantities
\begin{eqnarray}
&&
\Pi_{00}\equiv\Pi_{00}(i\xi,k_{\bot},0,\Delta,\mu),
\nonumber \\
&&
\Pi\equiv\Pi(i\xi,k_{\bot},0,\Delta,\mu).
\label{eq23}
\end{eqnarray}
\noindent
They can be obtained using Eqs.~(\ref{eq13}), (\ref{eq14}), and (\ref{eq16})
by performing integration with respect to $u$. The obtained results in the
cases $2\mu>\Delta$ and $2\mu\leqslant\Delta$ are different.
For the graphene sample used in our experiment the condition $2\mu>\Delta$
is satisfied (see Sec.~III) and after calculation one arrives at \cite{74}
\begin{eqnarray}
&&
\Pi_{00}=
\frac{8\alpha c\mu}{v_F^2}-\frac{\alpha\hbar k_{\bot}^2}{\tilde{q}}
\left\{
2M{\rm Im}(y_{\Delta,\mu}\sqrt{1+y_{\Delta,\mu}^2})
\right.
\nonumber \\
&&~~\left.
+(2-M)\left[2{\rm Im}\ln(y_{\Delta,\mu}+\sqrt{1+y_{\Delta,\mu}^2})
-\pi\right]\right\},
\nonumber \\[-1.5mm]
&& \label{eq24}\\
&&
\Pi=
-\frac{8\alpha \xi^2\mu}{cv_F^2}+2{\alpha\hbar\tilde{q} k_{\bot}^2}
\left[\vphantom{\frac{\pi}{2}}
M{\rm Im}(y_{\Delta,\mu}\sqrt{1+y_{\Delta,\mu}^2})
\right.
\nonumber \\
&&~~\left.
-(2-M){\rm Im}\ln(y_{\Delta,\mu}+\sqrt{1+y_{\Delta,\mu}^2})
+\frac{\pi}{2}(2-M)\right].
\nonumber
\end{eqnarray}
\noindent
Here,  the following notations are introduced
\begin{equation}
M=1+D^2, \quad
y_{\Delta,\mu}=\frac{\hbar\xi+2i\mu}{\hbar v_Fk_{\bot}\sqrt{M}}.
\label{eq25}
\end{equation}
\noindent
In the opposite case $2\mu\leqslant\Delta$ one has \cite{74}
\begin{equation}
\Pi_{00}=\Pi_{00}^{(0)},\qquad \Pi=\Pi^{(0)},
\label{eq25a}
\end{equation}
\noindent
where $\Pi_{00}^{(0)}$ and $\Pi^{(0)}$ are defined in Eq.~(\ref{eq14}) with
a replacement of $\xi_l$ for $\xi$.

We postpone a comparison between theoretical predictions of the above theory and
the measurement data to Sec.~VI. In the next section we consider the relative
magnitudes of the thermal correction and its constituents in the total Casimir
interaction for both real and pristine graphene samples and provide a qualitative
discussion of the origin and physical nature of the unusually big thermal
effect arising in graphene systems at short separations.

%%%%%%%%%%%%%%%%%%%%%%%%%%%%%%%%%
\section{Physical nature and magnitude of the thermal effect in real and
pristine graphene samples}

Let us calculate the relative thermal correction to the gradient of the Casimir
force acting between an Au-coated sphere of radius $R=60.35~\mu$m and a
graphene-coated SiO$_2$ plate which is given by
\begin{equation}
\delta_TF^{\prime}(a,T)=
\frac{\Delta_TF^{\prime}(a,T)}{F^{\prime}(a,0)},
\label{eq26}
\end{equation}
\noindent
where the absolute thermal correction $\Delta_TF^{\prime}$ is defined in Eq.~(\ref{eq21}).
We consider the graphene sheet with the experimental parameters $\mu =0.24~$eV and
$\Delta=0.29~$eV. Computations of the quantity $F^{\prime}(a,T)$ are performed by
Eqs.~(\ref{eq7}), (\ref{eq8}), (\ref{eq12}), (\ref{eq18}), and (\ref{eq19})
and of the quantity $F^{\prime}(a,0)$ --- by
Eqs.~(\ref{eq7}), (\ref{eq8}), (\ref{eq12}), (\ref{eq22}), and (\ref{eq24}).

%%%%%%%%%%%%%%%%%%%%%%__Fig._6__%%%%%%%%%%%%%%%%%%
\begin{figure}[!t]
\vspace*{-5.cm}
\centerline{\hspace*{0.5cm}
\includegraphics[width=5.50in]{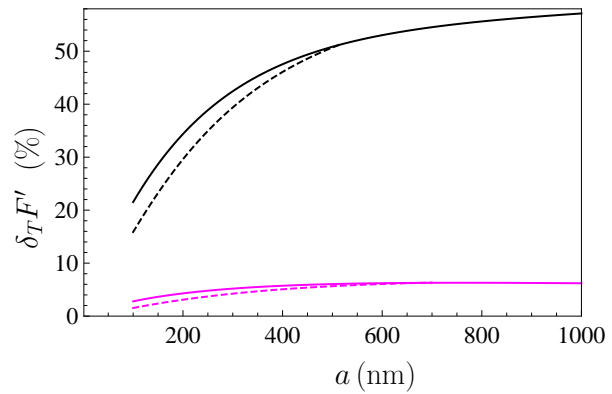}}
\vspace*{-9.2cm}
\caption{\label{fg6}   The computational results for the relative thermal
correction to the gradient of the Casimir force are shown as
functions of separation by the bottom and top pairs of lines
for an Au-coated sphere interacting either with
real graphene sheet deposited on a SiO$_2$ plate
or with freestanding real graphene sheet in vacuum,
respectively. The solid and dashed lines in each pair are
computed including and neglecting the explicit
dependence of the polarization tensor on temperature as a
parameter, respectively.
}
\end{figure}
%%%%%%%%%%%%%%%%%%%%%%%%%%%%%%%%%%%%%%%%%%%%%%%%%%%
The computational results for $\delta_TF^{\prime}$ at $T=294~$K as a function of
separation are presented by the bottom solid line in Fig.~\ref{fg6}.
As is seen in Fig.~\ref{fg6}, at separations of $a=100$, 200, 300, and 400~nm
the relative thermal correction in the experimental configuration reaches
2.79\%, 4.29\%, 5.19\%, and 5.73\% of the force gradient at $T=0$, respectively.
This effect is similar in magnitude to that one predicted by the Lifshitz theory
for the Casimir interaction between metallic test bodies described with inclusion
of the dissipation of conduction electrons. As discussed in Sec.~I, for metals
this prediction was excluded by the results of many experiments.

The bottom dashed line in Fig.~\ref{fg6} shows the computational results for
$\delta_TF^{\prime}$ under a condition that the quantity $F^{\prime}(a,T)$
is computed using the polarization tensor taken at $T=0$. This means that the
thermal correction in this case is implicit, i.e., fully determined by a summation
over the Matsubara frequencies, whereas an explicit dependence of the polarization
tensor on temperature as a parameter is disregarded.
As is seen from the bottom dashed line in Fig.~\ref{fg6},
at separations of $a=100$, 200, 300, and 400~nm
the implicit thermal correction is equal to
1.53\%, 3.10\%, 4.24\%, and 5.06\%, respectively.
Thus, with increasing separation the role of explicit dependence of the polarization
tensor on temperature gradually decreases and becomes negligibly small at
$a\geqslant700~$nm.

To determine the role of a SiO$_2$ substrate in the above results, we repeat
computations of the relative thermal correction $\delta_TF^{\prime}$ for the
configuration of an Au sphere and a freestanding graphene sheet preserving
unchanged all other
parameters of the experimental configuration. The computational results are shown
by the top pair of solid and dashed lines having the same meaning as the respective
lines of the bottom pair.
As is seen from the top solid line in Fig.~\ref{fg6},
at separations of $a=100$, 200, 300, and 400~nm
the thermal correction (\ref{eq26}) reaches much larger values of
21.5\%, 34.4\%, 42.4\%, and 47.5\%, respectively.
This means that in the absence of a substrate the thermal effect inherent to the
graphene sheet manifests itself more vividly.
The top dashed line in Fig.~\ref{fg6} illustrates the contribution of an implicit
thermal effect due to a summation over the Matsubara frequencies to the total
thermal correction in the case of a freestanding graphene sheet interacting with an
Au-coated sphere.
As is seen in Fig.~\ref{fg6}, at separations of $a=100$, 200, 300, and 400~nm
the implicit thermal effect contributes
15.9\%, 29.6\%, 39.4\%, and 46.1\% of the force gradient at $T=0$.
At separations exceeding 500~nm an explicit dependence of the polarization tensor
on $T$ does not lead to a noticeable contribution to the thermal correction (\ref{eq26}).

Up to this point we have considered the graphene sample with nonzero $\Delta$ and $\mu$
used in our experiment. It should be noted that for a pristine graphene possessing
$\Delta=\mu=0$ the thermal effect in the Casimir interaction at short separation
distances would be even much larger. To illustrate this, in Fig.~\ref{fg7} the
computational results for $\delta_TF^{\prime}$ in the
configuration of an Au-coated sphere interacting with a freestanding pristine
graphene sheet are shown by the top pair of solid and dashed lines as functions of
separation (the total and implicit thermal corrections, respectively).
As is seen  in Fig.~\ref{fg7},
at separations of $a=100$, 200, 300, 400, 700, and 1000~nm
the thermal correction  $\delta_TF^{\prime}$ defined in Eq.~(\ref{eq26}) is equal to
53.7\%, 115.5\%, 179.8\%, 245.6\%, 447.1\%, and 659.9\%, respectively.
According to Fig.~\ref{fg7}, for  pristine graphene the implicit thermal correction
plays a smaller role than for the experimental graphene sample. Thus,
at  $a=100$, 200, 300, 400, 700, and 1000~nm it is equal to
22.5\%, 61.1\%, 104.3\%, 149.8\%, 292.5\%, and 439.9\%, respectively.
As a result, the explicit thermal dependence of the polarization tensor contributes
31.2\%, 54.4\%, 75.5\%, 95.8\%, 154.6\%, and 212.0\%
of $F^{\prime}(a,0)$ at the same respective separations and does not disappear
when the separation increases.
%%%%%%%%%%%%%%%%%%%%%%__Fig._7__%%%%%%%%%%%%%%%%%%
\begin{figure}[!b]
\vspace*{-5.5cm}
\centerline{\hspace*{0.5cm}
\includegraphics[width=5.50in]{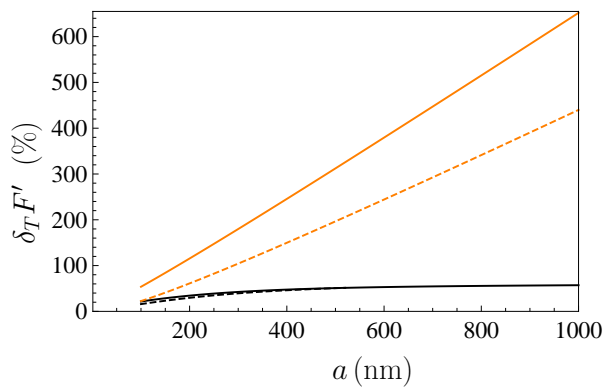}}
\vspace*{-9.2cm}
\caption{\label{fg7}   The computational results for the relative thermal
correction to the gradient of the Casimir force are shown as
functions of separation by the bottom and top pairs of lines
for an Au-coated sphere interacting with the freestanding either
real or pristine graphene sheet in vacuum, respectively.
The solid and dashed lines in each pair are computed
including and neglecting the explicit dependence of
the polarization tensor on temperature as a parameter,
respectively.}
\end{figure}
%%%%%%%%%%%%%%%%%%%%%%%%%%%%%%%%%%%%%%%%%%%%%%%%%%%

For comparison purposes, the bottom pair of solid and dashed lines in Fig.~\ref{fg7}
reproduces the top pair of lines in Fig.~\ref{fg6} related to the case of an Au-coated
sphere interacting with a freestanding real graphene sheet used in our experiment
(with $\Delta=0.29~$eV and $\mu=0.24~$eV). From Fig.~\ref{fg7} it is seen that
the replacement of a real with a pristine graphene sheet leads to a qualitatively
large increase of the thermal correction to the gradient of the Casimir force at
short separations.

{}From the above results it is seen that even the presence of a graphene sheet
deposited on a substrate significantly increases the thermal effect in the Casimir
interaction at short separations which constitutes only a small fraction of percent
for both metallic and dielectric materials. Because of this, it is interesting to
consider the so-called thermal regime of the Casimir interaction in the presence
 of graphene which takes place under the condition
 \begin{equation}
 F^{\prime}(a,0)\ll\Delta_TF^{\prime}(a,T),
 \label{eq27}
 \end{equation}
 \noindent
 i.e., when the thermal correction (\ref{eq21}) determines the major part of the
 force gradient at temperature $T$. This is the case when the term of the Lifshitz
 formula (\ref{eq7}) with $l=0$ becomes much larger than the sum of all remaining
 terms with $l\geqslant 1$.

In order to determine the role of graphene in reaching the thermal regime of the
Casimir interaction, we have computed separation distances between the parallel plates
made of different materials such that at larger separations the Casimir pressure
given by the zero-frequency term of the Lifshitz formula contributes more than 90\%,
95\%, and 99\% of the total Casimir pressure. The following cases were considered:
an Au plate and a SiO$_2$ plate; an Au plate and a SiO$_2$ plate coated with real
graphene sheet used in our experiment; an Au plate and a real graphene sheet;
an Au plate and a pristine graphene sheet; two pristine graphene sheets.
The obtained computational results are presented in Table~\ref{tb1} [we recall
that according to PFA the Casimir pressure between two parallel plates
$P=-F^{\prime}/(2\pi R)$, i.e., is proportional to the gradient of the Casimir
force in sphere-plate geometry used in our experiment].
%%%%%%___Table_I___%%%%%%%%%%%%%%%%%%%
\begingroup
\squeezetable
\begin{table}[b]
%\begin{table*}[b]
\caption{\label{tb1}
Separation distances in $\mu$m where the zero-frequency term of the Lifshitz
formula for the case of two parallel plates made of materials indicated in column~1
contributes more than 90\% (column~2), 95\% (column~3), and 99\% (column~4) of the total
Casimir pressure at larger separations. }
\begin{ruledtabular}
\begin{tabular}{cccc}
Plate &\multicolumn{3}{c}{$a~(\mu\mbox{m})$} \\
\cline{2-4}\\[-4mm]
materials & 90\% of $P$ & 95\% of $P$ & 99\% of $P$ \\ \hline
SiO$_2$---Au & 3.6 & 4.2 & 5.5 \\
(SiO$_2$+real graphene)---Au & 3.1 & 3.7 & 5.0 \\
(real graphene)---Au & 0.8 & 1.3 & 2.7 \\
(pristine graphene)---Au & 0.7 & 1.15 & 2.5 \\
(pristine graphene)---(pristine graphene) & 0.11 & 0.17 & 0.38
\end{tabular}
\end{ruledtabular}
%\end{table*}
\end{table}
\endgroup
%%%%%%%%%%%%%

As is seen in Table~\ref{tb1}, the presence of  a graphene sheet significantly decreases
the minimum separation distance from which the Casimir interaction is going into the
thermal regime. The thermal regime starts at especially short separations in the absence of a material substrate and for the pristine graphene sheets. Thus, for SiO$_2$--Au plates
the full thermal regime (99\% of the Casimir pressure) is reached only at $a\geqslant 5.5~\mu$m,
whereas for two pristine graphene sheets it is achieved at $a\geqslant 0.38~\mu$m.

In the end of this section we present a qualitative discussion of the physical reasons why
for two pristine graphene sheets the thermal regime starts at so short separations.
It is common knowledge that for ordinary materials the thermal regime starts at separations $a$
satisfying the condition \cite{32}
\begin{equation}
\frac{1}{2a}\ll\frac{\xi_1}{c}=2\pi\frac{k_BT}{\hbar c}.
\label{eq28}
\end{equation}
\noindent
This condition can be rewritten as
\begin{equation}
k_BT\gg\frac{1}{2\pi}k_BT_{\rm eff},\quad
k_BT_{\rm eff}\equiv\frac{\hbar c}{2a},
\label{eq29}
\end{equation}
\noindent
where $T_{\rm eff}$ is the so-called effective temperature.
Thus, according to numerical computations in Table~\ref{tb1}, for two plates made of
Au and SiO$_2$ the full thermal regime is reached at $a=5.5~\mu$m which corresponds
to the effective temperature $T_{\rm eff}\approx208.3~$K. In doing so
$T_{\rm eff}/(2\pi)\approx 33.2~$K so that at room temperature the inequality (\ref{eq29})
is well satisfied. Because of this, the thermal regime is also called the high-temperature
limit. In the high-temperature limit, the Casimir pressure determined by all Matsubara
frequencies with $l\geqslant 1$ is of the order of \cite{32}
\begin{equation}
P_{l\geqslant 1}(a,T)\sim \exp\left(-2\pi\frac{T}{T_{\rm eff}}\right),
\label{eq30}
\end{equation}
\noindent
i.e., is exponentially small.

For graphene, however, the situation is more complicated because the reflection
coefficients do not have the standard Fresnel form  (\ref{eq8}) but depend on the
polarization tensor. The major contribution to the Casimir pressure between two
graphene sheets is given by the TM polarization. From Eq.~(\ref{eq7}) we have
\begin{eqnarray}
&&
P_{\rm gr}(a,T)\approx -\frac{k_BT}{\pi} \sum_{l=0}^{\infty}{\vphantom{\sum}}^{\prime}
\int_0^{\infty}\!\!q_lk_{\bot}dk_{\bot}
\nonumber\\
&&~~~~~~~~~~
\times
\frac{r_{\rm TM,gr}^2(i\xi_l,k_{\bot})e^{-2aq_l}}{1-r_{\rm TM,gr}^2(i\xi_l,k_{\bot})e^{-2aq_l}},
\label{eq31}
\end{eqnarray}
\noindent
where the reflection coefficient on a freestanding graphene sheet $r_{\rm TM,gr}$
is obtained from $R_{\rm TM}$ defined in Eq.~(\ref{eq12}) by putting the dielectric
permittivity of a substrate $\varepsilon_l^{(2)}$ equal to unity
\begin{equation}
r_{\rm TM,gr}(i\xi_l,k_{\bot})=\frac{q_l\Pi_{00,l}}{2\hbar k_{\bot}^2+q_l\Pi_{00,l}}.
\label{eq32}
\end{equation}

To understand the qualitative physical reasons why graphene has a large thermal effect
already at relatively short separations, we restrict ourselves to the polarization tensor
taken at $T=0$ but calculated at the imaginary Matsubara frequencies (an account of the
explicit temperature dependence may only increase the thermal effect). Then, from
Eqs.~(\ref{eq14}) and (\ref{eq25a}) one finds
\begin{equation}
\Pi_{00,l}=\frac{\pi\alpha\hbar k_{\bot}^2}{\tilde{q}_l},
\label{eq33}
\end{equation}
\noindent
where we have taken into account that for a pristine graphene in accordance with Eq.~(\ref{eq15})
it holds $\Psi(0)=\pi$.

Substituting Eq.~(\ref{eq33}) in  Eq.~(\ref{eq32}), we obtain
\begin{equation}
r_{\rm TM,gr}(i\xi_l,k_{\bot})=\frac{\pi\alpha q_l}{\pi\alpha q_l+2\tilde{q}_l},
\label{eq34}
\end{equation}
\noindent
where $q_l$ is defined below Eq.~(\ref{eq7}) and $\tilde{q}_l$ in Eq.~(\ref{eq15}).

In the term of the Lifshitz formula (\ref{eq31}) with $l=0$, the reflection coefficient
(\ref{eq34}) takes the value
\begin{equation}
r_{\rm TM,gr}(0,k_{\bot})=\frac{\pi\alpha }{\pi\alpha +2\frac{v_F}{c}}
\approx 0.775.
\label{eq35}
\end{equation}

The distinctive feature of graphene is that the reflection coefficient (\ref{eq34})
depends on both $c$ and $v_F$. Because of this one can consider the region of separations
where
\begin{equation}
\frac{v_F}{2a}\ll\xi_1\ll\frac{c}{2a}.
\label{eq36}
\end{equation}
\noindent
The latter of these two inequalities is just the opposite to the condition (\ref{eq28})
required for reaching the
thermal regime between ordinary materials.
However, under the inequalities (\ref{eq36}) the reflection coefficient (\ref{eq34}) with
$l=1$ can be approximately presented in the form
\begin{equation}
r_{\rm TM,gr}(i\xi_1,k_{\bot})=
\frac{\pi\alpha }{2a(\frac{\pi\alpha}{2a} +2\frac{\xi_1}{c})}
=\frac{\pi\alpha }{\pi\alpha +\frac{4a\xi_1}{c}}.
\label{eq37}
\end{equation}
\noindent
Here, we used that the major contribution to Eq.~(\ref{eq31}) is given by
$k_{\bot}\approx 1/(2a)$ and that Eq.~(\ref{eq36}) results in
$q_1\approx 1/(2a)$ and $\tilde{q}_1\approx \xi_1/c$.
Taking into account that according to Eq.~(\ref{eq36}) $v_F\ll 2a\xi_1$, one
concludes from Eqs.~(\ref{eq35}) and (\ref{eq37}) that
\begin{equation}
r_{\rm TM,gr}(i\xi_1,k_{\bot})<r_{\rm TM,gr}(0,k_{\bot}).
\label{eq38}
\end{equation}
\noindent
The left-hand side of this inequality further decreases if $\xi_1$ is replaced
for $\xi_l$ with $l>1$. Thus, under the
condition (\ref{eq36}) a contribution of the zero-frequency term to Eq.~(\ref{eq31})
may become dominant in accordance to the results of numerical computations.

The first inequality in Eq.~(\ref{eq36}) can be identically rewritten in the form
\begin{equation}
k_BT\gg\frac{1}{2\pi}k_BT_{\rm eff}^{\rm gr},\quad
k_BT_{\rm eff}^{\rm gr}\equiv\frac{\hbar v_F}{2a},
\label{eq39}
\end{equation}
\noindent
which is similar to Eq.~(\ref{eq29}). Thus, for graphene, in addition to the standard
effective temperature $T_{\rm eff}$ defined  in Eq.~(\ref{eq29}), there exists one more
effective temperature $T_{\rm eff}^{\rm gr}$ defined  in Eq.~(\ref{eq39}) which is
lower than $T_{\rm eff}$ by a factor of 300. According to Ref.~\cite{40}, the
big thermal effect in the Casimir interaction between two graphene sheets at short
separations is controlled by the effective temperature $T_{\rm eff}^{\rm gr}$.
Our computational results and above
qualitative estimations show that the thermal regime of the Casimir interaction in graphene
systems is governed by two effective temperatures  $T_{\rm eff}^{\rm gr}$ and
$T_{\rm eff}$. In doing so at short separations
the thermal regime is determined by the much lower
temperature $T_{\rm eff}^{\rm gr}$.

%%%%%%%%%%%%%%%%%%%%%%%%%%%%%%%%%
\section{Comparison between experiment and theory}

The gradients of the Casimir force $F^{\prime}(a,T)$ between an Au-coated
sphere of $R=60.35 \pm 0.05~\mu$m radius and a graphene-coated
SiO$_2$ substrate at $T=294.0 \pm 0.5~$K temperature were computed by
Eqs.~(\ref{eq7}), (\ref{eq8}), (\ref{eq12}), (\ref{eq18}), and (\ref{eq19})
using the experimental values of the energy gap $\Delta=2.9 \pm 0.05~$eV
and chemical potential $\mu=0.24 \pm 0.01~$eV (see Sec. III).

It is well known that the Casimir interaction is influenced by
roughness on the interacting surfaces \cite{32,33,94,95,100a}. In the case
of small stochastic roughness with the rms amplitudes
$\delta_s=0.9 \pm 0.1~$nm and $\delta_g=1.5\pm 0.1~$nm on the sphere
and graphene surfaces, respectively (see Sec. II), it can be taken into account
multiplicatively \cite{32,33} resulting in the final expression for
the gradient of theoretical Casimir force
\begin{equation}
F_{\rm theor}^{\prime}(a,T)=\left(1+10\frac{\delta_s^2+\delta_g^2}{a^2}\right)
F^{\prime}(a,T).
\label{eq40}
\end{equation}

This expression was used to compute the upper and lower boundaries of
the top theoretical bands in Fig.~\ref{fg2} presenting allowed values of the
Casimir force gradient at $T=294~$K. These boundaries were computed in
the following most conservative way taking a proper account of all
errors which are present in the parameters used.

Thus, the upper boundary lines of the theoretical bands were calculated
with the largest allowed value of the chemical potential $\mu=0.25~$eV
and the smallest allowed value of the energy gap $\Delta=0.24~$eV.
This is explained by the fact that an increase of $\mu$ with fixed
$\Delta$ leads to a larger $F^{\prime}$ whereas an increase of $\Delta$
at a constant $\mu$ results in a smaller $F^{\prime}$ \cite{74}. The
obtained theoretical bands for $F_{\rm theor}^{\prime}$ were widened
to take into account the errors in the sphere radius and the 0.5\%
error in the force gradients arising from uncertainties in the
optical data of Au and SiO$_2$ (an error in the laboratory temperature
indicated above does not influence on the obtained results).

The theoretical bands for $F_{\rm theor}^{\prime}$ were also widened
to take into account small errors of PFA used in Eq.~(\ref{eq7}).
As it was shown in the literature \cite{96,97,98,99,100}, when using
the PFA one obtains slightly increased force gradients as compared
to the exact computational results in the sphere-plate geometry.
Because of this, we did not correct the upper lines of the theoretical
bands for the PFA error but introduced the maximum possible correction
factor of $(1-a/R)$ to the lower boundary lines.

As is seen in Fig.~\ref{fg2}, the upper theoretical bands computed at
$T=294~$K
are in a very good agreement with the measured gradients of the
Casimir force indicated as crosses over the entire measurements range
from 250 to 700~nm. The question arises as to whether the measurement data
demonstrate the presence of an unusually big thermal effect in the
Casimir force from graphene at short separations which is
considered in Sec.~V.

To answer this question, we have computed the gradients of the Casimir
force, $F^{\prime}(a,0)$, at zero temperature by using
Eqs.~(\ref{eq7}), (\ref{eq8}), (\ref{eq12}), (\ref{eq22}), and (\ref{eq24})
for the same parameters of the experimental configuration indicated
above. The obtained values of $F^{\prime}(a,0)$ were substituted to
Eq.~(\ref{eq40}) and the values of $F_{\rm theor}^{\prime}(a,0)$ were
calculated. The latter were used to find the upper and lower boundaries
of the theoretical bands for the
Casimir force gradient at $T=0$ in the same conservative way as
described above in the case of $T=294~$K. The results of this
calculation are presented by the bottom bands in Fig.~\ref{fg2}. As is seen
in Fig.~\ref{fg2}, the bottom theoretical bands computed at $T=0$ are more
narrow than the top ones computed at $T=294~$K. This is because our
graphene sample possesses a relatively large value of $\mu=0.24~$eV.
Calculations show that for such large values of the chemical
potential an impact of the energy gap on the polarization tensor
(and, as a consequence, on the reflection coefficients and force
gradients) at $T=0$ is considerably suppressed, as compared to the
case of $T=294~$K.

{}From Fig.~\ref{fg2} it is seen that the measurement data exclude the
theoretical predictions at $T=0$ shown by the bottom bands over
the wide separation region from 250 to 517~nm and, thus,
demonstrate the thermal effect in the Casimir interaction
arising from our graphene sample.

For a more illustrative demonstration of the observed thermal
effect, we also employ another way of comparison between experiment
and theory based on a consideration of differences between the
theoretical and mean experimental force gradients \cite{20,22,30,32,33}
\begin{equation}
F_{\rm theor}^{\prime}(a_i,\tilde{T})-F_{\rm expt}^{\prime}(a_i,T),
\label{eq41}
\end{equation}
where the experimental force gradients are given by the centers of the
crosses in Fig.~\ref{fg2} and the theoretical ones are computed with a step of
1 nm as described above.

In Fig.~\ref{fg8}, we plot the quantity (\ref{eq41}) as a function of
separation by the top and bottom sets of dots obtained at
$\tilde{T}=T=294~$K and $\tilde{T}=0~$K, $T=294~$K, respectively. The
confidence bands for the quantities (\ref{eq41}) found at
$\tilde{T}=T=294~$K (solid lines) and $\tilde{T}=0~$K, $T=294~$K (dashed
lines), respectively, take into account both the theoretical and
experimental errors determined at the 67\% confidence level. Note that
in addition to the theoretical errors considered previously, now we
also take into account an error arising from the fact that the
quantities $F_{\rm theor}^{\prime}(a_i,\tilde{T})$ in (\ref{eq41}) are
computed not over some separation region but at the experimental
separations $a_i$ determined with an error $\Delta a_i$. The bands
shown by the solid and dashed lines are slightly different because,
as discussed above, for our graphene sample an error in the energy
gap leads to different errors in the force gradients at $T=0$ and at
$T=294~$K.
%%%%%%%%%%%%%%%%%%%%%%__Fig._8__%%%%%%%%%%%%%%%%%%
\begin{figure}[!t]
\vspace*{-2cm}
\centerline{\hspace*{0.5cm}
\includegraphics[width=5.50in]{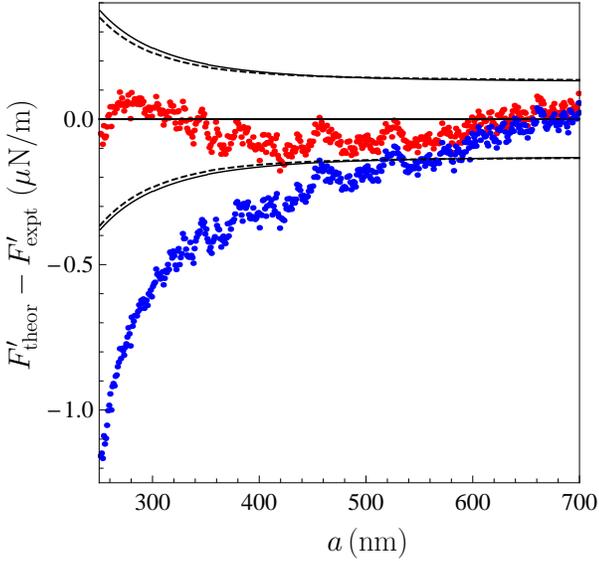}}
\vspace*{-9.2cm}
\caption{\label{fg8}  The differences between theoretical gradients of the
Casimir force computed either at $T=294~$K (top set of dots) or
at $T=0~$K (bottom set of dots) and mean experimental gradients
are shown as functions of separation. The solid and dashed lines
indicate the borders of the confidence intervals for the top and
bottom sets of dots, respectively, found at the 67\% confidence
level.   }
\end{figure}
%%%%%%%%%%%%%%%%%%%%%%%%%%%%%%%%%%%%%%%%%%%%%%%%%%%

As is seen in Fig.~\ref{fg8}, within the entire range of separations from
250 to 700~nm the top set of dots found at $T=294~$K is inside the
confidence band shown by the solid lines. This means that the
theoretical gradients of the Casimir force computed at $T=294~$K
are consistent with the measurement data. At the same time, the
bottom set of dots found at $T=0~$K is outside the confidence band
shown by the dashed lines over the wide range of separations from
250 to 517~nm, i.e., the theoretical results computed at $T=0$ are
experimentally excluded. These conclusions are in agreement with
those obtained above based on Fig.~\ref{fg2}.

The differences between the measurement data at $T=294~$K and computed at $T=0$
force gradients can be used to plot the thermal correction
$\Delta_{T}F^{\prime}$ defined in Eq.~(\ref{eq21}). In Fig.~\ref{fg9} it is
shown by dots as a function of separation in the region where the
theory at $T=0$ is experimentally excluded and cannot be used for
interpretation of the measurement data. The values of
$\Delta_{T}F^{\prime}$ at different separations shown in Fig.~\ref{fg9} are
consistent with the theoretical values of $\delta_{T}F^{\prime}$
computed in Sec.~V for our graphene sample. This can be easily
verified by using the computational results for the gradients of
the Casimir force presented in Fig.~\ref{fg2}. Thus, the performed experiment
demonstrates an unusual thermal effect in graphene systems which
becomes noticeable even at relatively short separations of a few
hundred nanometers.
%%%%%%%%%%%%%%%%%%%%%%__Fig._9__%%%%%%%%%%%%%%%%%%
\begin{figure}[!t]
\vspace*{-4.cm}
\centerline{\hspace*{0.7cm}
\includegraphics[width=5.50in]{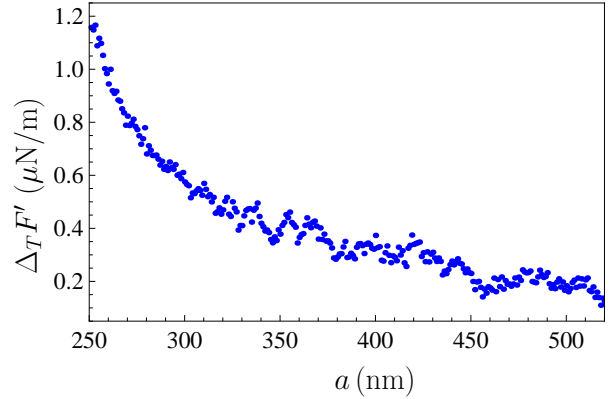}}
\vspace*{-9.2cm}
\caption{\label{fg9}  The thermal correction to the gradient of the Casimir
force found as a difference between the mean experimental gradients
measured at $T=294~$K and the theoretical ones computed at $T=0~$K
is shown by dots as a function of separation. }
\end{figure}
%%%%%%%%%%%%%%%%%%%%%%%%%%%%%%%%%%%%%%%%%%%%%%%%%%%

%%%%%%%%%%%%%%%%%%%%%%%%%%%%%%%%%
\section{Conclusions and discussion}

In the foregoing, we have described measurements of the thermal
Casimir interaction between an Au-coated sphere and a
graphene-coated substrate performed at $T$=294K in high vacuum
by means of a custom built AFM cantilever
based setup operated in the dynamic regime.
Using the two sets of measurements each of which contains 21
experimental runs, we have obtained the mean gradients of the
Casimir force and determined their random, systematic, and total
experimental errors over the separation region between a sphere
and a graphene sheet from 250 to 700 nm. In doing so all the
experimental parameters, including the absolute separations,
and their errors were determined by means of electrostatic
calibration. For the substrate supporting the graphene sheet,
a sufficiently thick fused silica plate has been used which,
as was proposed in Ref. \cite{69}, should make it possible to
observe the unusual thermal effect in graphene systems at short
separations predicted in Ref. \cite{40}.

According to the literature provided in Sec. I, an experimental
discovery of this effect for graphene would be of great
fundamental importance because a similar effect had long been
predicted by the Lifshitz theory for metals described by the
conventional Drude response function \cite{38}, but was
experimentally excluded by numerous precision experiments
\cite{19,20,21,22,23,24,25,26,27,28,29,30,31}.

We have performed measurements of the energy gap and impurity
concentration in the graphene sample used and compared the
experimental mean gradients of the Casimir force with theory
based on the first principles of quantum electrodynamics at
nonzero temperature with no fitting parameters. For this
purpose, the response function of graphene was described by
the polarization tensor at nonzero temperature depending on
the energy gap and chemical potential which is found in the
framework of the Dirac model. The experimental results are
 in a very good agreement with the theoretical ones
computed at $T=294~$K over the entire measurement range within
the limits of experimental and theoretical errors. The
theoretical gradients of the Casimir force computed using
the same theory with the same experimental parameters at
$T=0~$K are conclusively excluded by the measurement data
over the separation region from 250 to 517 nm. Thus, the
presence of an unusual thermal effect in graphene systems
at short separations is confirmed experimentally.

We have investigated the dependence of the thermal correction
to the gradient of the Casimir force between a sphere and a
graphene sample taking into account the values of the
energy gap and chemical
potential of graphene and also  the presence of a substrate.
The case of two parallel freestanding sheets of a pristine
graphene, originally studied in Ref. \cite{40}, was also
considered. It was confirmed that an observed size of the
thermal effect is in agreement with that for a pristine
graphene taking into account respective decrease due to the
presence of a substrate and nonzero values of the energy
gap and chemical potential of the graphene sheet used.

An experimental verification of the thermal effect, which
is observed in the Casimir interaction with graphene at
short separations, offers a clearer view on why a similar
effect is experimentally excluded for metallic test
bodies. The key point is that for graphene the response function
to quantum fluctuations is determined theoretically on the
basis of first physical principles. It is nonlocal (i.e.,
depends both on the frequency and on the wave vector) and
pertains equally to quantum fluctuations on the mass shell
(the propagating waves) and off the mass shell (the evanescent
waves). Then, it is reasonable that the theoretical
predictions for the Casimir interaction, which has contributions
from both the propagating and evanescent waves, obtained using
this formalism are confirmed experimentally.
In comparison for metals,
their response functions are found partially experimentally
from tabulated values \cite{36}
and partially using the theoretical extrapolation both given
by the effect of only
 propagating waves. These response functions are reliably
tested only in the area of quantum fluctuations on the mass
shell. Note that it is even impossible to experimentally
test the transverse components of their spatially nonlocal
generalizations in the off-mass-shell area \cite{37}.
This may be the reason why the Lifshitz theory using the
 standard Drude model or its generalization for the case
of frequency-dependent relaxation parameter (the so-called
Gurzhi model) fails to predict the correct values of the Casimir
force between metallic test bodies \cite{101}.

Thus, information obtained from using graphene
leads us to conclude that the Casimir puzzle for metals could
be resolved by making the spatially nonlocal modification of
the Drude model in the area of evanescent waves which leaves
the response to the propagating waves almost unchanged. Such
an attempt was already undertaken in Ref.~\cite{102}. The
suggested spatially nonlocal Drude-like response functions
take into account the dissipation properties of conduction
electrons, as does the standard Drude model, and
simultaneously bring the Lifshitz theory in agreement with
measurements of the Casimir force between metallic surfaces.
It is pertinent to note that the Lifshitz theory using the
nonlocal Drude-like response functions introduced in
Ref.~\cite{102} satisfies the Nernst heat theorem both for
metals with perfect crystal lattices and for lattices with
the structural defects  \cite{103} (we recall that the
Casimir entropy calculated using the standard Drude model
violates this fundamental theorem for metals with perfect
crystal lattices \cite{32,33,35}).

To conclude, the observation of an unusual thermal effect
in graphene systems at short separations, reported in this
paper, may stimulate resolution of several other fundamental
problems and also be useful for numerous applications of
graphene in physics and nanotechnology.

%%%%%%%%%%%%%%%%%%%%%%%%%%%%%%%%%
\section*{Acknowledgments}
The work of M.~L., Y.~Z.~and U.~M.~was partially supported by the NSF
grant PHY-2012201.
The work of G.~L.~K. and V.~M.~M. was partially supported by the Peter
the Great Saint Petersburg Polytechnic
University in the framework of the Russian state assignment for basic
research (project N FSEG-2020-0024).
V.~M.~M.~was partially funded by the Russian Foundation for Basic
Research, Grant No. 19-02-00453 A. His work was also partially
supported by the Russian Government Program of Competitive Growth
of Kazan Federal University.
The authors thank Jianlin Liu, Yongtao Cui, Xiong Xu, Yuan Li, and
Yanwei He Huang of the University of California, Riverside for help with
and discussions regarding STS.

%%%%%%%%%%%%%%%%%%%%%%%%%%%%%%%%

%%%%%%
\end{document}